\documentclass[11pt,twocolumn]{IEEEtran}
\usepackage{caption2}
\captionstyle{hang}

\usepackage{stfloats}

\usepackage{amsmath,mathrsfs,amssymb,epsfig,psfrag,amsthm,bm,multirow, graphicx,color}
\usepackage{ mathrsfs,dsfont,yfonts }
\usepackage{bbold}
\usepackage{textcomp}

\IEEEoverridecommandlockouts

\newtheorem{proposition}{Proposition}

\newtheorem{lemma}{Lemma}

\begin{document}

\title{Downlink Scheduling over Markovian Fading Channels}

\author{\emph{Wenzhuo Ouyang, Atilla Eryilmaz, and Ness B.
Shroff}
\vspace{-9pt}

\vspace{-5pt}
\thanks{Wenzhuo Ouyang and Atilla Eryilmaz are with the Department of ECE, The Ohio State University (e-mails: ouyangw@ece.osu.edu, eryilmaz@ece.osu.edu).
Ness B. Shroff holds a joint appointment in both the Department of ECE and the Department of CSE at The Ohio State University (e-mail: shroff@ece.osu.edu). }
\thanks{A preliminary version of this paper appeared in INFOCOM 2012.}
\thanks{This work was supported in part by NSF grants CAREER-CNS-0953515, CCF-0916664, CNS-1012700, DTRA grant HDTRA 1-08-1-0016, Qatar National Research Fund (QNRF) under the National Priorities Research Program (NPRP) grant NPRP 09-1168-2-455, and ARO MURI award  W911NF-08-1-0238.}
}
\maketitle

\begin{abstract}
We consider the scheduling problem in downlink wireless
networks with heterogeneous, Markov-modulated, ON/OFF channels. It
is well-known that the performance of scheduling over fading
channels relies heavily on the accuracy of the available Channel
State Information (CSI), which is costly to acquire. Thus, we
consider the CSI acquisition via a practical ARQ-based feedback
mechanism whereby channel states are revealed at the end of only scheduled
users' transmissions. In the assumed presence of
temporally-correlated channel evolutions, the desired scheduler must
optimally balance the \emph{exploitation-exploration trade-off},
whereby it schedules transmissions both to exploit those channels
with up-to-date CSI and to explore the current state of those with
outdated CSI.

In earlier works, Whittle's Index Policy had been suggested as a
low-complexity and high-performance solution to this problem.
However, analyzing its performance in the typical scenario of
statistically heterogeneous channel state processes has remained
elusive and challenging, mainly because of the highly-coupled and
complex dynamics it possesses. In this work, we overcome these
difficulties to rigorously establish the asymptotic optimality
properties of Whittle's Index Policy in the limiting regime of
many users. More specifically: (1) we prove the \emph{local
optimality} of Whittle's Index Policy, provided that the
initial state of the system is within a certain neighborhood of a
carefully selected state; (2) we then establish the \emph{global
optimality} of Whittle's Index Policy under a recurrence assumption
that is verified numerically for our problem. These results
establish that Whittle's Index Policy possesses analytically provable optimality
characteristics for scheduling over heterogeneous and
temporally-correlated channels.
\end{abstract}

\section{Introduction}

Channel fluctuation is an intrinsic characteristic of wireless communications. Such a variation calls for allocation of the wireless resources in a dynamic manner, leading to the classic \emph{opportunistic scheduling principle} (e.g., \cite{Knopp}\cite{JSAC_Liu}). Under the assumption that the instantaneous channel state information (CSI) is fully available to the scheduler, many efficient opportunistic scheduling algorithms (e.g., \cite{tassiulas}-\cite{Atilla}) have been proposed and extensively studied.

More recent works have focused on designing scheduling algorithms under imperfect CSI, where the channel state is modeled as independent and identically distributed (\textit{i.i.d.}) processes across time (e.g., \cite{2stage}-\cite{Allerton}). On the other hand, although the \textit{i.i.d.} channel model brings ease of analysis, it fails to capture the time-correlation of the fading channels \cite{Tse}. Specifically, it fails to exploit the channel memory, which is a critical resource for making scheduling decisions. However, designing efficient scheduling schemes under time-correlated channels with imperfect CSI is a very challenging problem. The challenge is mainly because of the difficulty in making the classic `exploitation versus exploration' trade-off  (e.g., \cite{clinical,reinforce}), in which a scheduler needs to strike a balance between selecting the channels with up-to-date channel memory that guarantees high immediate gains, or to explore the channels with outdated CSI for more informed decisions and associated future throughput gains.


We consider the downlink scheduling problem where a base station transmits to the users within its transmission range, subject to scheduling constraints. To model the time correlations present over fading channels, we assume that wireless channels evolve as Markov-modulated ON/OFF processes. The channel state information is obtained from ARQ-based feedback, only \emph{after} each scheduled transmission. Nevertheless, due to time correlation, the memory of the past channel state can be used to predict the current channel state \emph{prior to} scheduling decision.
Hence, channel memory should be intelligently exploited by the scheduler in order to achieve high throughput performance.

In a related work \cite{YingShakkottai}, a similar problem is
considered under delayed CSI, where it is assumed that
perfect CSI is available within a maximum delay, which is in turn smaller than the delay experienced by the ARQ feedback used for collision detection. These assumptions allow  the scheduling decisions to be
decoupled from CSI acquisition, which leads to the
development of centralized as well as distributed
schedulers. However, this approach does not use ARQ as a means of acquiring improved channel quality information. In contrast, in our setup the nature of ARQ feedback creates an implicit impact of scheduling
decisions on the CSI feedback, which completely transforms the nature of
the optimal scheduler design, and therefore requires a different
approach. Under the scenario where all the channels have
\emph{identical Markov statistics},
round-robin-based algorithms (e.g., \cite{Liu}-\cite{Neely_utility}) have
been shown to possess optimality properties in throughput performance.
However, the round-robin-based algorithms are no longer optimal in
\emph{asymmetric scenarios}, e.g., when different channels have different
Markov transition statistics, as is naturally the case in typical
heterogeneous conditions.


Under the asymmetric scenarios, our downlink scheduling problem is an example of the classic Restless Multiarmed Bandit Problem
(RMBP) \cite{Whittle}. Low-complexity Whittle's Index Policies\hspace{3pt}\cite{Whittle}\hspace{3pt}for the downlink scheduling problem have been
proposed in \cite{Zhao_index}\cite{Infocom11} based on RMBP theory.
However, although Whittle's Index Policy can bring significant
throughput gains by exploiting the channel memory \cite{Infocom11},
the analytical characterization of its performance under asymmetric
scenarios is very challenging and prohibitively technical. This is because asymmetry leads to a sophisticated interplay of memory evolution among channels with heterogeneous characteristics, which brings a significant challenge to the analysis of Whittle's Index Policy not present in the perfectly symmetric scenario. 


For RMBP problems under general scenarios, Whittle's Index Policy has been proven in \cite{Weber} to be asymptotically optimal
as the number of users grows, provided a non-trivial condition,
known as Weber's condition, holds. Nonetheless, Weber's condition
concerns the global convergence of a non-linear differential
equation, which is extremely difficult to verify even
numerically in our downlink scheduling scenario. In \cite{ApproxRMBP}, optimality properties of general RMBP is studied, where a sub-optimal BALANCED-INDEX policy, as well as a THRESHOLD-WHITTLE policy, are proved to provide $2-$approximation performance, i.e., achieves at least half of the optimal reward. Our work takes a different approach than \cite{ApproxRMBP} to specifically study the per-user throughput performance of the Whittle's Index Policy for downlink scheduling, and consider the strict optimality metric in the asymptotic regime when the number of users scales. 

In this paper, we take significant steps in analyzing the optimality
properties of Whittle's Index Policy for the downlink scheduling
problem in the presence of channel heterogeneity. Specifically, our contributions are as follows.

\begin{itemize}
\item We apply the Whittle's index framework to our downlink scheduling problem and identify the optimal policy for the problem with a relaxed constraint in Section~\ref{sec:bound}. This policy, with carefully selected randomization, provides a performance upper bound to Whittle's Index Policy.

\item We establish the local optimality of Whittle's Index Policy in the asymptotic regime when the number of users scales in Section~\ref{sec:local}. Specifically, we show that the performance of the index policy can get arbitrarily close to that of the relaxed-constraint optimal policy, provided that the initial state of the system is within a certain neighborhood of a carefully selected state.

\item Based on the local optimality result, under a numerically verifiable recurrence assumption, we then establish the global optimality of Whittle's Index Policy in the limiting regime of many users in Section~\ref{sec:global}.
\end{itemize}

\section{System Model and Problem Formulation}
\label{sec:model}

\subsection{Downlink Wireless Channel Model}

We consider a time-slotted, wireless downlink system with one base
station and $N$ users. The wireless channel $C_i[t]$ between base
station and user $i$ remains static within each time slot $t$ and
evolves stochastically across time slots, independently across
users. We adopt the simplest non-trivial model of time-correlated
fading channels by considering two-state ON/OFF channels, where the
state space of channel $i$ is $\mathcal{\bm S}_i=\{0, 1\}$, with
the value of each state representing the transmission rate a channel
can support at the state.
\begin{figure}
\centering
\includegraphics[width=3in]{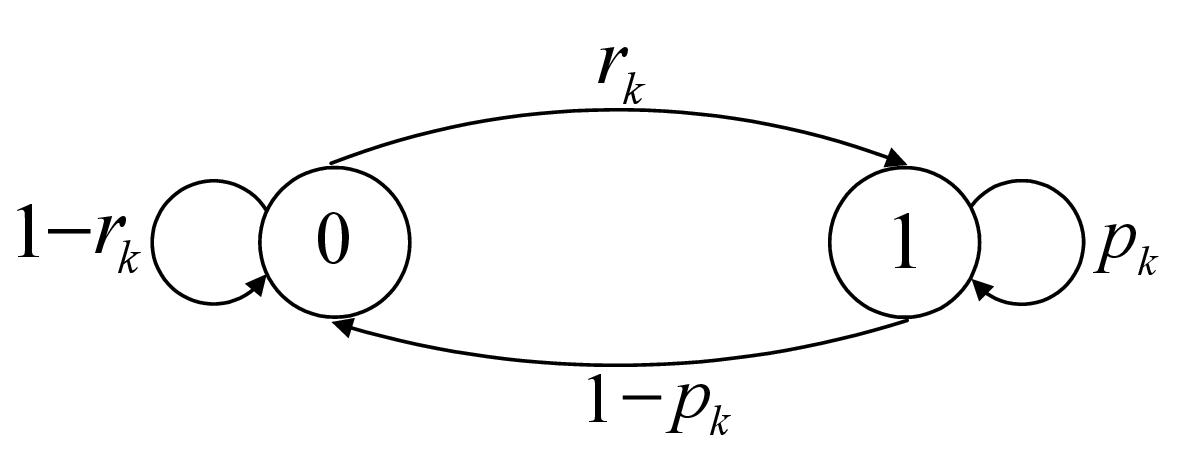}
\caption{Two state Markov chain model for channels in class $k$.}
\label{fig:chain}
\end{figure}

One important component of our model is the inclusion of channel heterogeneity that the users will typically experience in real systems. Such asymmetry creates a significant challenge to the design and analysis of optimal scheduling schemes compared to perfectly symmetric channels. To avoid cumbersome notation and unessential technical complications, in this work we model channel asymmetry by considering only \emph{two classes} of channel statistics. Specifically, for all the channels in class $k$, $k{=}1,2$, their states evolve according to the same Markov statistics. However, these characteristics differ between classes. The state transition of channels in class $k$ is depicted in Fig.~\ref{fig:chain}, represented by a $2\times 2$ probability transition matrix,
\vspace{-4pt}
\begin{align}
\mathbb{\bm P}_k=\begin{bmatrix}
p_k&1-p_k\\
r_k&1-r_k\\
\end{bmatrix},\nonumber
\end{align}

\vspace{-7pt}
\noindent where

\vspace{-18pt}
\begin{align}
p_k&:= \textrm{prob$\big(C_i[t]{=}1 \ \big | \ C_i[t{-}1]{=}1\big)$,}\nonumber \\
r_k&:= \textrm{prob$\big(C_i[t]{=}1 \ \big | \ C_i[t{-}1]{=}0 \big)$.}\nonumber
\end{align}

\vspace{-4pt}
\noindent for channel $i$ in class $k$. The number of class $k$ channels is $\gamma_k N$, $k\in \{1,2\}$ with $\gamma_k$ being the \emph{proportion} of channels in class $k$ with respective to the total number $N$ of channels.

We study the scenario where all the Markovian channels are positively correlated, i.e., $p_k > r_k$ for $k{=}1,2$. This assumption, which is commonly made in this domain (e.g., \cite{Neely_capacity, Neely_utility, sugu_aslm}), means that the channel evolution has a positive auto-correlation. Hence, roughly speaking, the channel has a stronger potential to stay in its previous state than jumping to another, which is typical especially in slow fading environment. For ease of exposition, we shall exclude the trivial case when $r_k\hspace{1pt}{=}\hspace{1pt}0$ or $p_k\hspace{1pt}{=}\hspace{1pt}1$, $k=1,2$. \vspace{-5pt}

\subsection{Scheduling Model -- Belief Value Evolution}

We assume that the base station can simultaneously transmit to at most $\alpha N \hspace{1pt}{\in}\hspace{1pt}\mathbb{Z}^+$ users in a time slot without interference, where $\alpha \hspace{1pt}{\in}\hspace{1pt}(0,1]$ stands for the maximum \emph{fraction} of users that can be activated. For example, in a multi-channel communication model, $\alpha$ would correspond to the fraction of all users that can be simultaneously serviced in unit time. However, the scheduler does not know the exact channel state in the current slot when the scheduling decision is made. Instead, the scheduler maintains a \emph{belief value} $\pi_i[t]$ for each channel $i$, which is defined as the probability of channel $i$ being in the ON state at the beginning of slot $t$. The accurate channel state is revealed via ACK/NACK feedback from the scheduled users, only at the end of each time slot after the data is transmitted. This accurate channel state feedback is in turn used by the scheduler to update the belief values.

For user $i$ in class $k$, $k{=}1,2$, let $a_i[t] {\in} \{0,1\}$ indicate whether the user is selected for transmission in slot $t$. Then, from the definition the belief values, $\pi_i[t]$ evolves as follows,

\begin{align}
\label{eq:evolve}
\hspace{-3pt}\pi_i[t{+}1]{=}\hspace{-2pt}
\begin{cases}
p_k,& \text{if $a_i[t]{=}1$, $C_i[t]{=}1$,}\\
r_k,& \text{if $a_i[t]{=}1$, $C_i[t]{=} 0$,}\\
\pi_i[t] p_k{+}(1{-}\pi_i[t])r_k, & \text{if $a_i[t]{=}0.$}
\end{cases}
\end{align}

In our setup, belief values are known to be sufficient statistics to represent the past scheduling decisions and feedback (e.g., \cite{Javidi,Sondik_thesis}). In the meanwhile, in our ON/OFF channel model, $\pi_i[t]$ also equals to the expected throughput contributed by channel $i$ if it is scheduled in time slot $t$.

For a user in class $k$, $k{=}1,2$, we use $b^k_{c,l}$ to denote its
belief value when the most recent observed channel was
$c\in \{0,1\}$, and is $l$ slots in the past. From the belief
update rule (\ref{eq:evolve}), $b^k_{c,l}$ can be calculated as a
function of $l {\geq} 1$ as,
\begin{align}
b^k_{0,l}{=}\frac{r_k\hspace{1pt}{-}\hspace{1pt}(p_k-r_k)^l
r_k}{1+r_k-p_k}, \
b^k_{1,l}{=}\frac{r_k\hspace{1pt}{+}\hspace{1pt}(1-p_k)(p_k-r_k)^l}{1+r_k-p_k}.
\nonumber
\end{align}

\begin{figure}
\centering
\includegraphics[width=3.3in]{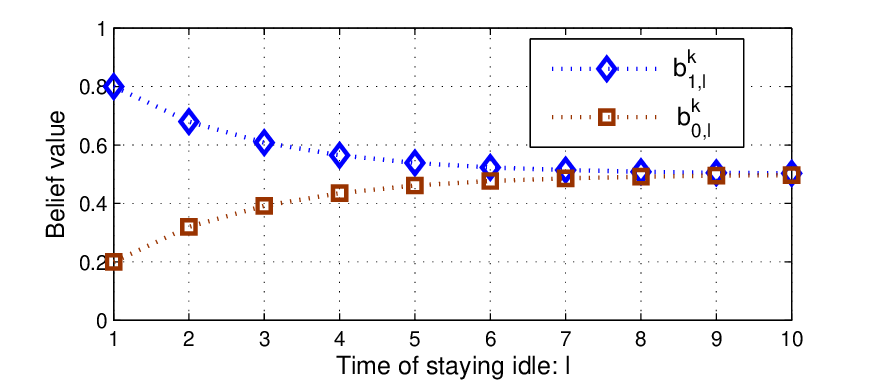}
\caption{Belief values update when staying idle, $p_k=0.8$, $r_k=0.2$, $b^k_s=0.5$.}
\label{fig:Qupdate}
\end{figure}

Fig.~\ref{fig:Qupdate} illustrates the belief value update when a channel stays idle
(i.e., $a_i{=}0$). It is clear
that if the scheduler is never updated of the state of channel $i$
(in class $k$), the belief value will converge to its stationary
probability of being ON, denoted by the stationary belief value
$b_s^{k}{:=}{r_k}/{(1{+}r_k{-}p_k)}$.

The vector $\vec{\bm \pi}[t]{=} ( \pi_1[t], {\cdots}, \pi_N[t])$ denotes the belief values of all channels at the beginning of slot $t$. We use $\mathcal{B}_k$ to represent the set of the belief values for class $k$ channels, where $\mathcal{B}_k{=}\{b^k_{s}, b^k_{c,l}, c \hspace{1pt}{\in}\hspace{1pt} \{0,1\}, l\hspace{1pt}{\in}\hspace{1pt} \mathbb{Z}^+ \}$. We assume that the system starts to operate from slot $t=0$. At the beginning of slot $0$, for each channel the scheduler has either observed its channel state before, or has never been updated of its channel state, i.e., with belief value $b^k_{s}$. It is then clear that, based on the belief update rule (\ref{eq:evolve}), $\pi_i[t] \in \mathcal{B}_k$ for all $t\geq0$, i.e., each belief value $\pi_i[t]$ evolves over countably many states.

In the rest of the paper, we shall use `belief value' and `belief state' interchangeably.

\subsection{Downlink Scheduling Problem -- POMDP Formulation}
We consider the broad class $U$ of (possibly non-stationary) scheduling policies that makes a scheduling decision based on the history of observed channel states and scheduling actions.
The downlink scheduling problem is then to identify a policy in $U$ that maximizes the infinite horizon, \emph{time average expected throughput}, subject to the constraint on the number of users selected at each time slot. Given the initial state $\vec{\bm \pi}[0]$, the problem is formulated as,
\begin{align}
\max_{u \in U}& \hspace{5pt} \liminf_{T\rightarrow \infty} \frac{1}{T} E\Big[\sum_{t=0}^{T-1} \sum_{i=1}^N \pi_i[t] \cdot a^u_i[t] \Big | \vec{\bm \pi}[0] \Big] \label{eq:strgt_obj} \\
s.t.& \hspace{0.1in}  \sum_{i=1}^{N} a^u_i[t]\leq \alpha N , \quad \forall t. \label{eq:strgt}
\end{align}
where the belief value $\pi_i[t]$ evolves according to rule (\ref{eq:evolve}) based on the scheduling decision $a^u_i[t]$ under policy $u$. Such an objective is standard in literature for Markov Decision Processes under the long term average reward criteria (e.g., \cite{Eitan}). Noting that since the scheduling decisions are made based on incomplete knowledge of channel states, this problem is a Partially Observable Markov Decision Process \cite{Sondik_thesis}.

This problem is in fact an example of Restless Multiarmed Bandit
Problem (RMBP) \cite{Whittle}. For a general RMBP, finding an
optimal solution is PSPACE-hard \cite{Tsitsiklis}. However, for the
downlink scheduling problem at hand, a low-complexity Whittle's Index Policy was proposed in \cite{Zhao_index}\cite{Infocom11} based on the RMBP
theory that inherently exploits the channel memory when making scheduling
decisions. For detailed descriptions of general RMBP
and Whittle's Index Policy for downlink scheduling, please refer to
\cite{Whittle}-\cite{Infocom11}.

For the downlink scheduling problem, we note that there is only limited analytical characterization of Whittle's Index Policy, which is restricted in perfectly symmetric scenarios where Whittle's Index Policy takes a special round-robin form \cite{Zhao_index}. In asymmetric cases, however, the scheduling decision no longer takes the form of round-robin, bringing sophisticated complications in belief value evolutions that are tightly coupled among channels, which significantly complicates the analysis. The main focus of this paper is to analytically characterize the performance of Whittle's Index Policy in the asymmetric case with two classes of channels.

\section{Upper Bound on Achievable Throughput}
\label{sec:bound}

We begin our analysis by characterizing an upper bound to the throughput performance of all feasible downlink scheduling policies that satisfies the constraint (\ref{eq:strgt}). The upper bound is obtained from a fictitious policy which is optimal for the downlink scheduling problem under a \emph{relaxed constraint}.

Note here that such relaxation is also a crucial step in the study
of the general RMBP problem. Yet, our analysis, being specific to
the downlink scheduling problem, has its novelties, as we shall
remark on later.

\subsection{Average-Constrained Relaxed Scheduling Problem}

We consider an associated relaxed problem of (2)-(3) that only requires an \emph{average number} of users to be activated in the long run, defined as follows
\begin{align}
\max_{u \in U}& \hspace{5pt} \liminf_{T\rightarrow \infty} \frac{1}{T} E\Big[\sum_{t=0}^{T-1} \sum_{i=1}^N \pi_i[t] \cdot a^u_i[t] \Big | \vec{\bm \pi}[0] \Big] \label{eq:relaxed_obj} \\
s.t.& \hspace{6pt}  \limsup_{T\rightarrow \infty} \frac{1}{T}E\Big[\sum_{t=0}^{T-1} \sum_{i=1}^{N} a^u_i[t] \Big ]\leq \alpha N . \label{eq:relaxed}
\end{align}


Note that, contrary to the stringent constraint (\ref{eq:strgt}), the relaxed constraint (\ref{eq:relaxed}) allows the activation of more than $\alpha$ fraction of users in each time slot, provided the long term average fraction does not exceed $\alpha$. Hence the optimal policy under this relaxed constraint, which we shall identify next, provides a throughput upper bound to any policy that satisfies the stringent constraint.

\subsection{Optimal Policy for the Relaxed Problem}

We remark that the relaxed problem is also an important component of Whittle's analysis of general RMBPs \cite{Whittle}, in which an optimal policy for the relaxed problem is developed based on the \emph{Whittle's index values}. Following the approach of classic RMBP framework \cite{Whittle}, in our downlink scenario, we identify an optimal policy for the relaxed problem based on Whittle's indices.

Specifically, for channels in class $k$, the Whittle's index value
$W_k(\pi)$ is assigned to each belief state $\pi \in \mathcal{B}_k$.
These index values intuitively capture the exploitation and
exploration value to be gained from scheduling the associated
channel when its belief value is $\pi$.  This characteristic of
$W_k(\pi)$ is also illustrated in Section~\ref{sec:num:trade-off} via
numerical investigations. The index value function is expressed in closed form as 
\begin{align}
\label{eq:indices}
W_k(\pi){=}\hspace{-2pt}\begin{cases}
\frac{(b^k_{0,l}{-}b^k_{0,l+1}) (l{+}1){+}b^k_{0,l+1}}{1{-}p_k{+}(b^k_{0,l}{-}b^k_{0,l+1})l{+}b^k_{0,l+1}} &\text{if $r_k{\leq}\pi{=}b^k_{0,l}{<}b^k_s$} \\
\frac{r_k}{(1-p_k)(1+r_k-p_k)+r_k} &\text{if $b^k_s \leq \pi\leq p_k$}
\end{cases}
\end{align}

Note that the above expression is a modified version of the expression in \cite{Zhao_index}. Details of the derivation can be found in \cite{Wenzhuo_infocom12}.

The following two characteristics
they possess are primarily significant for our analysis:
\begin{itemize}
\item $W_k(\pi)$ monotonically increases with $\pi \in \mathcal{B}_k$.
\item $W_k(\pi)\in [0,1]$ for all $\pi \in \mathcal{B}_k$.
\end{itemize}

The next lemma identifies an index-based policy with \emph{appropriate randomization} that is optimal for the relaxed constraint problem. This policy schedules each user based on its own belief value, independently from other users. The proof of the Lemma can be found in \cite{Zhao_index}.

\begin{lemma}
\label{lemma:thres_relax}
For the problem under relaxed constraint, there exists an optimal stationary policy $\phi^*$, parameterized by the threshold $\omega^*$ and a randomization parameter $\rho^* {\in} (0,1]$, such that
\vspace{3pt}

\noindent(i) Channel $i$ in class $k$ is scheduled if
$W_k(\pi_i[t])\hspace{1pt}{>}\hspace{1pt}\omega^*$, and stays idle
if \hspace{3pt}$W_k(\pi_i[t])\hspace{1pt}{<}\hspace{1pt}\omega^*$.
If $W_k(\pi_i[t])\hspace{1pt}{=}\hspace{1pt}\omega^*$, it is
scheduled with probability $\rho^*$.

\vspace{3pt}

\noindent(ii) The parameters $\omega^*$ and $\rho^*$ are such that,
under policy $\phi^*$, the relaxed constraint (\ref{eq:relaxed}) is
strictly satisfied with equality.

\end{lemma}

From now on, we shall denote $\phi^*$ as the `\emph{Optimal Relaxed
Policy}'. {For technical purposes, we henceforth assume $\alpha$ is such that $\rho^*{\neq} 1$}. Since each $\alpha$ value maps to a unique $(\omega^*, \rho^*)$ pair \cite{Wenzhuo_wiopt}, only countably many $\alpha$ values correspond to $\rho^*{=} 1$, i.e., achieved by deterministic policies. Therefore, the set of $\alpha {\in} (0,1]$ for which $\rho^*{\neq} 1$ has Lebesgue measure one.

\subsection{Steady State Distribution of Belief Values}

We next present the transition structure of the belief values under
Optimal Relaxed Policy, captured in the following lemma. The
structure will be critical in the development of our subsequent
main results.

\vspace{2pt}

\begin{lemma}
\label{lemma:pos_rec} For each channel in class $k$, under the
Optimal Relaxed Policy, the structure of belief value evolution
depends on the threshold $\omega^*$ of policy. \vspace{2pt}

\noindent(i) If $\omega^*\hspace{1pt}{<}\hspace{1pt}W_k(b^k_s)$, then the belief value evolution of each class $k$ channels is positive recurrent with a finite recurrent class.
\vspace{2pt}

\noindent(ii) If $\omega^* \hspace{1pt}{\geq}\hspace{1pt}W_k(b^k_s)$, the belief value evolution is transient. With probability $1$, ultimately no channel in class $k$ will transmit.
\end{lemma}

\noindent \textbf{Proof:} The proof of this lemma follows from the
monotonic structure of belief evolution, as shown in
Fig.~\ref{fig:Qupdate}. Details are included in Appendix
\ref{appen:recur}. $\hfill \blacksquare$ \vspace{3pt}

Thus, if $\omega^* \hspace{1pt}{\geq}\hspace{1pt} \max\{ W_1(b^1_s),
W_2(b^2_s)\}$, the above analysis reveals that ultimately no user
transits, corresponding to the trivial case of $\alpha
N\hspace{1pt}{=}\hspace{1pt}0$. Also, if $\omega^*$ is between
$W_1(b^1_s)$ and $W_2(b^2_s)$, the class with the smaller
$W_k(b^k_s)$ will eventually transit into a passive mode, hence
reducing the system to a well-understood scenario with a single class of channels \cite{Liu}\cite{Javidi}. Thus, here we focus on the
heterogeneous case of $\omega^* \hspace{1pt}{<}\hspace{1pt}
W_k(b^k_s), k{=}1,2$, where the steady-state belief value distribution
exists for both classes under the Optimal Relaxed Policy.

\subsection{Upper bound on achievable throughput}
The throughput performance of Optimal Relaxed Policy provides an throughput
upper bound for all policies under the stringent constraint. The
value of such an upper bound clearly depends on the number of users
in each class $\gamma_k N$, $k\hspace{1pt}{=}\hspace{1pt}1,2$, as well as the fraction
$\alpha$ of users allowed for activation. Denoting $\bm
\gamma\hspace{1pt}{=}\hspace{1pt}[\gamma_1, \gamma_2]$, we represent the time average expected throughput of the Optimal Relaxed Policy as
$\upsilon^N(\bm \gamma, \alpha)$. The following lemma states
that, as long as $\bm \gamma$ and $\alpha$ are given, the \emph{per-user} throughput (i.e., $\upsilon^N(\bm \gamma, \alpha)/N$) is independent of $N$.

\begin{lemma}
\label{lemma:parameter}
Given $\bm \gamma$ and $\alpha$, $\frac{\upsilon^N(\bm \gamma, \alpha)}{N}$ is independent of $N$, denoted henceforth as $r(\bm \gamma,\alpha)$.
\end{lemma}

\noindent \textbf{Proof:} The proof follows from showing that, when the number of users $N$ grows, as long as the proportion of each class of channels stays the same and the fraction $\alpha$ of users activated does not change, the form of Optimal Relaxed Policy does not change. Since each user is scheduled independently, the throughput $\upsilon^N(\bm \gamma, \alpha)$ is proportional to $N$, establishing the lemma. Details are provided in Appendix~\ref{appen:peruser}. $\hfill \blacksquare$

\vspace{3pt}

We hence refer to the $(\bm \gamma, \alpha)$ pair as `\emph{system parameters}'. Therefore $N r(\bm \gamma, \alpha)$ provides a throughput upper bound to any policy in the same system under the stringent constraint (\ref{eq:strgt}). Equivalently, $r(\bm \gamma, \alpha)$ provides a per-user throughput performance upper bound to all policies that satisfies the stringent constraint.

We next describe Whittle's Index Policy for the strictly-constrained problem (\ref{eq:strgt_obj})-(\ref{eq:strgt}), and later study the closeness of its performance to the upper bound established here.


\section{Whittle's Index Policy Description}
\label{sec:num:index}

In this section we formally introduce Whittle's Index Policy for
solving the stringently-constrained downlink scheduling problem
(\ref{eq:strgt_obj})-(\ref{eq:strgt}).


\subsection{Whittle's Index Policy}
The Optimal Relaxed Policy, along with the Whittle's index values,
gives consistent ordering of belief values with respective to the
indices. For instance, under the Optimal Relaxed Policy, if it is
optimal to schedule one channel, it is then optimal to transmit to
other channels with higher index values. So the Whittle's index
value gives an intuitive order of how attractive the channel is for
scheduling. This intuition leads to Whittle's Index Policy
\cite{Zhao_index} under the stringent constraint on the maximum number of
channels that can be scheduled.

\vspace{3pt}
\noindent\textbf{Whittle's Index Policy:} \emph{At the beginning of each time slot, the channel $i$ in class $k$ is scheduled if its Whittle's index value $W_k(\pi_i)$ is within the top $\alpha N$ index values of all channels in that slot, with arbitrary tie-breaking while assuring a total $\alpha N$ channels being scheduled.}
\vspace{2pt}

Whittle's Index Policy is attractive because it has very low
complexity, and it was observed via numerical investigations to
yield significant throughput performance gains over the scheduling
strategies that does not utilize channel memory \cite{Infocom11}.
The main focus of our work is to analytically understand the approximate or asymptotic optimality of Whittle's Index Policy in asymmetric scenarios.

\subsection{Whittle's Index Policy over Truncated State Space}

Recall from Section~\ref{sec:model} that the belief values evolve over a countable
state space, also note that if a channel is not scheduled for a long time, its
belief value will get arbitrarily close to its stationary belief
value. This motivates us to consider a truncated version of the
belief value evolution whereby the belief value is set to its steady
state if the corresponding channel is not scheduled for a large
number, say $\tau$, slots. This mild assumption facilitates more
tractable performance analysis of the policy.  Thus,
if a class $k$ user is not scheduled for $\tau$ time slots, its channel state
history is entirely forgotten and its belief value will transit to
the stationary belief value $b_s^k$, where the truncation $\tau$ is assumed
to be very large.

Whittle's Index Policy is then implemented over the truncated
belief state, which differs from the non-truncated case merely in
the truncated belief value evolution. We believe that, the truncated
scenario can provide arbitrarily close approximation to the original
system when $\tau$ is large. More importantly, as we shall see in
the following two sections, Whittle's Index Policy, implemented
over the truncated belief state space, achieve asymptotically
optimal performance as long as the truncation is sufficiently large.

\section{Local Optimality of Whittle's Index policy}
\label{sec:local}

In this section, we study the optimality properties of Whittle's
Index Policy for downlink scheduling, over a large truncated belief
space. This result forms the basis for the subsequent global
optimality result in Section~\ref{sec:global}. We start by introducing a state
space over which the local optimality will be established.

\subsection{System State Vector}
\label{sec:local:Z}

We define the \emph{system state} $\bm {\bm Z}^N$ as a vector
that represents the proportion of channels in each belief value, over
the truncated space when the total number of users is $N$, i.e.,
${\bm Z}^N=\big[{\bm Z}^{\>1,N}, {\bm Z}^{\>2,N}\big]$,
with
\begin{align}
\nonumber {\bm Z}^{\>k,N}{=}[ Z_{0,1}^{k,N},{\cdots},
Z_{0,\tau}^{k,N}, Z_{s}^{k,N}, Z_{1,\tau}^{k,N},{\cdots},Z_{1,1}^{k,N} ], k{=}1,2.
\end{align}
where $Z^{k,N}_{c,l}$ and $Z^{k,N}_{s}$ respectively denote the
\emph{proportion} of channels in the corresponding belief state
$b^k_{c,l}$ and $b^k_{s}$, with respect to the total number of users
$N$. Hence, each element of ${\bm Z}^N$ is a multiple of $1/N$
so that ${\bm Z}^N$ takes values in a lattice with mesh size
$1/N$. Noting that the total number of users in each class does not
change over time, for any $N$ the system state ${\bm Z}^N[t] \in
\mathcal{Z}$ where
\begin{align}
\mathcal{Z}:{=}\{{\bm Z}^N \geq 0: Z^{k,N}_{s}{+}\sum_{c,l}
Z^{k,N}_{c,l}{=}\gamma_k, \hspace{2pt} k{=}1,2 \}. \label{eq:beta}
\end{align}

The system state vector ${\bm Z}^N[t]$ does not distinguish
users with the same belief state, thus its dimension will not scale
with $N$. Therefore, compared with $\vec{\bm \pi}[t]$, it provides a
more convenient representation of the system belief state.
Furthermore, ${\bm Z}^N[t]$ fully determines the instantaneous
throughput gain in slot $t$ under both Whittle's Index Policy
and the Optimal Relaxed Policy (introduced in
Lemma~\ref{lemma:thres_relax}), because the instantaneous
throughput gains under both policies are only determined by the
distribution of the channels with different belief values, not their
identities.


From Lemma~\ref{lemma:pos_rec} and the subsequent remarks,
under the operation of the Optimal Relaxed Policy, the belief state
evolution of each channel is positive recurrent with a steady-state
distribution. The following lemma also establishes the independence
of this steady-state distribution from $N$, and defines a useful
parameter for future use.




\begin{lemma}\label{lemma:zeta_invar}
Given the system parameters $(\bm \gamma, \alpha)$, the system state
vector ${\bm Z}^N[t]$ under the Optimal Relaxed Policy converges
in distribution to a random vector, denoted as ${\bm
Z}^N[\infty].$ The mean of ${\bm Z}^N[\infty]$ is
independent of $N$ and is denoted as

\vspace{-18pt}
\begin{align}
\vec{\bm \zeta}^{\alpha}_{\bm \gamma}{:=} E\big[{\bm
Z}^N[\infty]\big]. \nonumber
\end{align}
\end{lemma}

\noindent{\textbf{Proof:}} This lemma follows from a similar
principle to the one we established in Lemma~\ref{lemma:parameter}.
For details, please refer Appendix~\ref{appen:zeta_invar}.
\hspace{2in} $\hfill \blacksquare$

It is easy to see that $\vec{\bm \zeta}^{\alpha}_{\bm \gamma} \hspace{1pt}{\in}\hspace{1pt}
\mathcal{Z}$ and the form of $\vec{\bm \zeta}^{\alpha}_{\bm \gamma}$
fully determines the time average throughput of the Optimal Relaxed
Policy. Therefore, the vector $\vec{\bm \zeta}^{\alpha}_{\bm
\gamma}$ provides an important benchmark for our asymptotic
analysis. If, in the long run under Whittle's Index Policy, the system state ${\bm Z}^N[t]$ stays close to
$\vec{\bm \zeta}^{\alpha}_{\bm \gamma}$, it indicates that Whittle's Index Policy will have throughput performance close to
that of the Optimal Relaxed Policy -- the throughput upper bound. To
capture the closeness, we define the $\delta$ neighborhood of
$\vec{\bm \zeta}^{\alpha}_{\bm \gamma}$ as
\begin{align}
\label{eq:nbhd} \Omega_{\delta}(\vec{\bm \zeta}^{\alpha}_{\bm
\gamma})= \{{\bm Z} \in \mathcal{Z}: ||{\bm Z}- \vec{\bm
\zeta}^{\alpha}_{\bm \gamma}||\leq \delta \},
\end{align}
for $\delta>0$, where $||\cdot||$ stands for Euclidean distance. We
are now ready to state and prove our first main result regarding a
form of local optimality of Whittle's Index Policy.


\subsection{Local Optimality of Whittle's Index Policy}
Under the system parameters $(\bm \gamma, \alpha)$, we let
$R_{T}^{N}(\bm \gamma, \alpha, \bm x)$ represent the time average
throughput obtained over the time duration $0
\hspace{1pt}{\leq}\hspace{1pt} t \hspace{1pt}{<}\hspace{1pt} T$
under Whittle's Index Policy, conditioned on the initial system
state ${\bm Z}^N[0]=\bm x$, i.e.,
\begin{align}
\nonumber R_{T}^{N}(\bm \gamma, \alpha, \bm
x)\hspace{1pt}{:=}\hspace{1pt}\frac{1}{T}E \Big[\sum_{t=0}^{T-1}
\sum_{i=1}^{N} \pi_i[t] a_i^{ind}[t] \Big| {\bm Z}^N
[0]\hspace{1pt}{=}\hspace{1pt}\bm x \Big],
\end{align}
where $(a_i^{ind}[t])_i$ denotes the scheduling decision vector made
by Whittle's Index Policy at time $t.$

Recall from Lemma~\ref{lemma:parameter} that $r(\bm \gamma, \alpha)$
denotes the per-user throughput under the Optimal Relaxed Policy,
which serves as an upper bound on Whittle's Index Policy
performance. The next proposition characterizes the local
convergence property of Whittle's Index Policy performance to $r(\bm
\gamma, \alpha)$.

\begin{proposition}
\label{prop:local_conv} Under the system parameters $(\bm \gamma,
\alpha)$, there exists a $\delta >0$ neighborhood
$\Omega_{\delta}(\vec{\bm \zeta}^{\alpha}_{\bm \gamma})$ of
$\vec{\bm \zeta}^{\alpha}_{\bm \gamma}$ such that, if the initial
system state $\bm x$ is within $\Omega_{\delta}(\vec{\bm
\zeta}^{\alpha}_{\bm \gamma})$ , then
\begin{align}
\lim_{T \rightarrow \infty} \lim_{m \rightarrow \infty}  \frac{R_{T}^{N_m}(\bm \gamma, \alpha, \bm x)}{N_m}\hspace{1pt}{=}\hspace{1pt}r(\bm \gamma, \alpha). \nonumber
\end{align}
where $\{ N_m \}_m$ is any increasing sequence of positive integers
with $\alpha N_m$, $\gamma_k N_m \in \mathbb{Z}^+$, for $k=1,2$ and
all $m$.
\end{proposition}


\noindent \textbf{Proof Outline:} Here, we give a high level
description of the proof for an intuitive understanding, and refer the reader to
Appendix~\ref{appen:local} for the rigorous
derivation. \vspace{1pt}

$\bullet$ We start by defining a fluid approximation, whereby the
discrete-time evolution of ${ \bm Z}^N[t]$ under Whittle's Index
Policy is modeled as a deterministic vector ${ \bm z}[t] \in
\mathcal{Z}$ that evolves {in discrete time} over $\mathcal{Z}$ and is independent of $N.$ Under this fluid approximation, the users are no longer
unsplittable entities so that the state space of ${ \bm z}[t]$ is no
longer restricted to a lattice as it was for ${ \bm Z}^N[t]$. Also,
the fluid approximation ${ \bm z}[t]$ evolves in a deterministic
manner, in contrast to the stochastic transition of ${\bm Z}^N[t]$.
The evolution of ${\bm z}[t]$ is defined by a {difference} equation
as a function of the \emph{expected} state change of ${ \bm
Z}^N[t]$ under Whittle's Index Policy as follows
\begin{align}
\label{eq:fluid_sketch}{\bm z[t+1] {-} \bm z[t] \Big |_{{ \bm
z}[t]={ \bm z}} }\hspace{-3pt}{=} E\Big[{ \bm Z}^N[t+1]{-}{ \bm Z}^N[t] \Big| { \bm
Z}^N[t]{=}{ \bm z}\Big],
\end{align}
where $N$ is any integer for which ${\bm z}$ is a feasible state.

$\bullet$ We then establish local convergence of the fluid
approximation model when ${\bm z}[0]$ is within a small enough
$\delta$ neighborhood $\Omega_{\delta}(\vec{\bm \zeta}^{\alpha}_{\bm
\gamma})$ of $\vec{\bm \zeta}^{\alpha}_{\bm \gamma}$. We show the
convergence by first noting that the differential equation
(\ref{eq:fluid_sketch}) is linear within a wider convex region than
$\Omega_{\delta}(\vec{\bm \zeta}^{\alpha}_{\bm \gamma})$. Within
this region, we obtain a closed form expression of the right hand
side of (\ref{eq:fluid_sketch}), which enables us to investigate the
eigenvalue structure of the linear differential equation. We show
{that each eigenvalue $\lambda$ satisfies $|\lambda|< 1$} and apply standard linear
system theory to establish the local convergence.

$\bullet$ We then connect the fluid approximation model ${\bm z}[t]$
to the discrete-time stochastic system state ${\bm Z}^N[t]$ by using a discrete-time extension of
Kurtz's Theorem, which can be interpreted as an
extension of the strong law of large numbers to random processes
(see \cite{Weiss_LD}). Essentially, it states that, over any finite
time duration $[0,T]$, the actual
system evolution ${\bm Z}^N[t]$ can be made arbitrarily close
to the above fluid approximation ${\bm z}[t]$ by increasing the
number of channels $N$ sufficiently, {with exponential convergence rate}.

$\bullet$ The previous convergence result, together with the local convergence
result of the fluid evolution ${\bm z}[t]$ to $\vec{\bm
\zeta}^{\alpha}_{\bm \gamma}$, enables us to establish the local
convergence of the system state ${\bm Z}^N[t]$ to $\vec{\bm
\zeta}^{\alpha}_{\bm \gamma}$ as the number of users $N$ grows,
provided that the initial state ${\bm Z}^N[0] \in \Omega_{\delta}(\vec{\bm
\zeta}^{\alpha}_{\bm \gamma})$.
Hence the system state under Whittle's Index Policy will stay close
(in a probabilistic sense) to the expectation $\vec{\bm
\zeta}^{\alpha}_{\bm \gamma}$ of the system state under the Optimal
Relaxed Policy, which, in turn, indicates that the throughput
performance of Whittle's Index Policy will approach the throughput
upper bound $r(\bm \gamma, \alpha)$, as expressed in the
proposition.

We again emphasize that the technical details of the outlined steps
are fairly intricate and are moved to Appendix~\ref{appen:local}. $\hfill \blacksquare$ \vspace{5pt}


Proposition~\ref{prop:local_conv} illustrates an interesting local
optimality property of Whittle's Index Policy as the number of users $N$
and the time horizon $T$ increases while the system parameters
$(\bm \gamma, \alpha)$ stay the same. It indicates that, under
Whittle's Index Policy, as long as the initial state ${\bm Z}^N[0]$
is close enough to $\vec{\bm \zeta}^{\alpha}_{\gamma}$, the average
per-user throughput over any finite time duration will get
arbitrarily close to the Optimal Relaxed Policy performance as the
number of users scales. \vspace{3pt}


\noindent \textbf{Remark: } We note that the sequence $\{N_m\}_m$ is
used to guarantee that the number of channels in each class, as well
as the number of scheduled users, take integer values. In fact, our
result can be generalized to all $N$ by slightly perturbing $\bm
\gamma$ and $\alpha$ as a function of $N$ but assuring their limits
are well-defined.

\section{Global Optimality of Whittle's Index Policy}
\label{sec:global}
The above local optimality result heavily relies on the initial
state ${\bm Z}^N[0]$ being close to $\vec{\bm \zeta}^{\alpha}_{\bm
\gamma}$, which is difficult to guarantee. In this section, we study
the global optimality of the infinite horizon throughput performance
of Whittle's Index Policy starting from any initial state. We begin
our analysis by presenting the recurrence structure of the system state.

\begin{lemma}
\label{lemma:recur}
Under system parameters $(\bm \gamma, \alpha)$, for any $\epsilon>0$, if the number of users $N$ is large enough, 

\noindent(i) The system state ${\bm Z}^N[t]$ evolves as an aperiodic
Markov chain, in a state space that contains only one recurrent
class.

\noindent(ii) There exists at least one recurrent state within the
$\epsilon$ neighborhood $\Omega_{\epsilon}(\vec{\bm
\zeta}^{\alpha}_{\bm \gamma})$ of $\vec{\bm \zeta}^{\alpha}_{\bm
\gamma}$.
\end{lemma}

\noindent \textbf{Proof:}  We prove this lemma by constructing probability paths from any
state to the neighborhood $\Omega_{\epsilon}(\vec{\bm
\zeta}^{\alpha}_{\bm \gamma})$. Details of the proof are included in Appendix~\ref{appen:recur}. $\hfill \blacksquare$
\vspace{4pt}


This lemma states that ${\bm Z}^N[t]$ will ultimately enter any
small neighborhood of $\vec{\bm \zeta}^{\alpha}_{\bm \gamma}$ when
$N$ is large enough. Together with
Proposition~\ref{prop:local_conv}, this result shows promise for
establishing the global asymptotic optimality of Whittle's Index
Policy. This is plausible because once ${\bm Z}^N[t]$ enters
$\Omega_{\delta}(\vec{\bm \zeta}^{\alpha}_{\bm \gamma})$, the
performance of Whittle's Index Policy \emph{afterwards} can
get very close to its upper bound as $N$ scales, as established in
Proposition~\ref{prop:local_conv}. However, since we consider the infinite horizon time average throughput, this argument would break down if the time it takes for ${\bm Z}^N[t]$ to enter $\Omega_{\delta}(\vec{\bm \zeta}^{\alpha}_{\bm \gamma})$ also scales
up with $N$. This observation
motivates us to introduce a useful assumption, which will later be justified
(in Section~\ref{sec:num:just}) via numerical studies.

\vspace{5pt}


\noindent \textbf{Assumption $\Psi$}: For each $\epsilon{>}0$, let
$\Gamma_{\bm x}^N(\epsilon)$ represent the first time of reaching
$\Omega_{\epsilon}(\vec{\bm \zeta}^{\alpha}_{\bm \gamma})$ starting
from ${\bm Z}^N[0]= {\bm x}$, i.e.,
\begin{align}
\Gamma_{\bm x}^N(\epsilon)=\min \{t: {\bm Z}^N[t] \in
\Omega_{\epsilon}(\vec{\bm \zeta}^{\alpha}_{\bm \gamma}) \big | {\bm
Z}^N[0]= {\bm x} \}. \nonumber
\end{align}
Then we assume that, the expected time of reaching
$\Omega_{\epsilon}(\vec{\bm \zeta}^{\alpha}_{\bm \gamma})$ is
bounded by a constant $M_{\epsilon}{<}\infty$, i.e.,
\begin{align}
E\big[\Gamma_{\bm x}^N
(\epsilon)\big]{\leq}M_{\epsilon},\nonumber
\end{align}
for all $\bm x$ and large enough $N$.

\vspace{7pt}

Since for each $N$, ${\bm Z}^N[t]$ under Whittle's Index Policy is
recurrent and aperiodic with a finite state space, there exists a
steady-state distribution associated with ${\bm Z}^N[t]$. As before,
we use ${\bm Z}^N[\infty]$ to denote the associated limiting random
vector. The next lemma establishes that, under Assumption $\Psi$, the
distribution of ${\bm Z}^N[\infty]$ approaches a point-mass at
$\vec{\bm \zeta}^{\alpha}_{\bm \gamma}$ as $N$ scales. Here, again,
the sequence $\{N_m\}_m$ is defined in the same way as in
Proposition \ref{prop:local_conv}.

\vspace{-2pt}
\begin{lemma}
\label{lemma:steady_dist} Under Assumption $\Psi$ and system
parameters $(\bm \gamma, \alpha)$, for any $\epsilon>0$, the steady
state probability of ${\bm Z}^N[t]$ under Whittle's Index Policy
satisfies
\vspace{-3pt}
\begin{align}
\lim_{m \rightarrow \infty} P\big({\bm Z}^{N_m}[\infty] \in
\Omega_{\epsilon}(\vec{\bm \zeta}^{\alpha}_{\bm \gamma})\big)=1.
\nonumber
\end{align}
\end{lemma}

\vspace{-2pt}
\noindent \textbf{Proof:} The proof utilizes Theorem $6.89$ from
\cite{Weiss_LD}, which builds on the following arguments.

\begin{figure}
\centering
\includegraphics[width=3.3in]{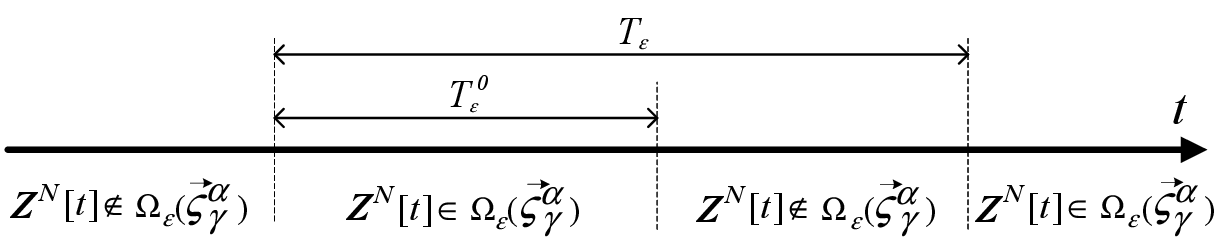}
\vspace{-8pt}
\caption{Transition behavior of ${\bm Z}^N[t]$ in steady state.}
\label{fig:invar_meas}
\vspace{-12pt}
\end{figure}

Note that $\epsilon>0$ can be selected to be small enough for the
following argument. As depicted in Fig.~\ref{fig:invar_meas}, we let
$T_{\epsilon}$ be a random variable denoting, in steady state, the
time duration between \emph{consecutive} hitting times into the
neighborhood $\Omega_{\epsilon}(\vec{\bm \zeta}^{\alpha}_{\bm
\gamma})$ from outside of the neighborhood. Let $T^0_{\epsilon}$
denote the time duration from the time ${\bm Z}^N[t]$ enters the
neighborhood $\Omega_{\epsilon}(\vec{\bm \zeta}^{\alpha}_{\bm
\gamma})$ from outside until the time it leaves. Hence, the
expected proportion of time that ${\bm Z}^N[t]$ stays outside this
neighborhood is
$(E[T_{\epsilon}]-E[T^0_{\epsilon}])/E[T_{\epsilon}$].

We know that the numerator $E[T_{\epsilon}]-E[T^0_{\epsilon}]$ is uniformly
bounded for all $N$ due to Assumption $\Psi$. However, as $N$
increases, it is more likely for ${\bm Z}^N[t]$ to stay within the
neighborhood for a long time before exiting it (based on the convergence of fluid approximation model and Kurtz's Theorem in the proof of Proposition~\ref{prop:local_conv}). Thus,
$E[T^0_{\epsilon}],$ and hence the denominator $E[T_{\epsilon}]$, grow to infinity
as $N$ scales. Therefore, the expected proportion of time spent
outside $\Omega_{\epsilon}(\vec{\bm \zeta}^{\alpha}_{\bm \gamma})$
vanishes as $N$ scales up, which leads to the statement of the
lemma. Details of the proof can be found in Appendix~\ref{appen:Invar_meas}. $\hfill \blacksquare$

\vspace{3pt} Under Whittle's Index Policy with system parameters
$(\bm \gamma, \alpha)$, we let $R^{N}_{\bm x}(\bm \gamma, \alpha)$ be
the achieved infinite horizon, time average throughput, conditioned
on the initial system state ${\bm Z}^N[0]{=}\hspace{1pt} \bm x$,
i.e.,
\begin{align}
\nonumber R^{N}_{\bm x}(\bm \gamma,
\alpha)\hspace{1pt}{:=}\hspace{1pt}\lim_{T\rightarrow \infty}
\frac{1}{T}E \Big[\sum_{t=0}^{T-1} \sum_{i=1}^{N} \pi_i[t]
a_i^{ind}[t] \Big|{\bm Z}^{N}[0]={\bm x} \Big].
\end{align}

\vspace{-5pt}From Lemma~\ref{lemma:steady_dist} we know that, in steady-state,
the system state $\bm Z^{N_m}[\infty]$ is increasingly
concentrated around $\vec{\bm \zeta}^{\alpha}_{\bm \gamma}$ as $m$
increases, regardless of the initial state ${\bm x}.$ We build on
this to establish the global asymptotical optimality of Whittle's
Index Policy.


\begin{proposition}
\label{prop:asymp} Under Assumption $\it \Psi$, for any initial
system state $\bm x$, we have

\vspace{-12pt}\begin{align}
\lim_{m \rightarrow \infty} \frac{R^{N_m}_{\bm x}(\bm \gamma, \alpha)}{N_m} = r(\bm \gamma, \alpha). \nonumber
\end{align}
Since $r(\bm \gamma, \alpha)$ is an upper bound on the maximum
achievable per-user throughput by any policy, this implies that
Whittle's Index Policy is optimal in the many user regime.
\end{proposition}

\noindent \textbf{Proof:} We prove this result by decomposing $R^{N}_{\bm x}(\bm \gamma,
\alpha)$ as a summation of the expected throughput conditioned on whether the system state is within or outside an arbitrarily small $\epsilon$ neighborhood of $\vec{\bm \zeta}^{\alpha}_{\bm \gamma}$. Since the latter has diminishing probability according to Lemma~\ref{lemma:steady_dist}, the expected throughput of Whittle's Index Policy can get arbitrarily close to that of Optimal Relaxed Policy. Details of the proof are provided in Appendix~\ref{appen:global}.$\hfill \blacksquare$ \vspace{3pt}

\begin{figure}
\centering
\includegraphics[width=3.8in]{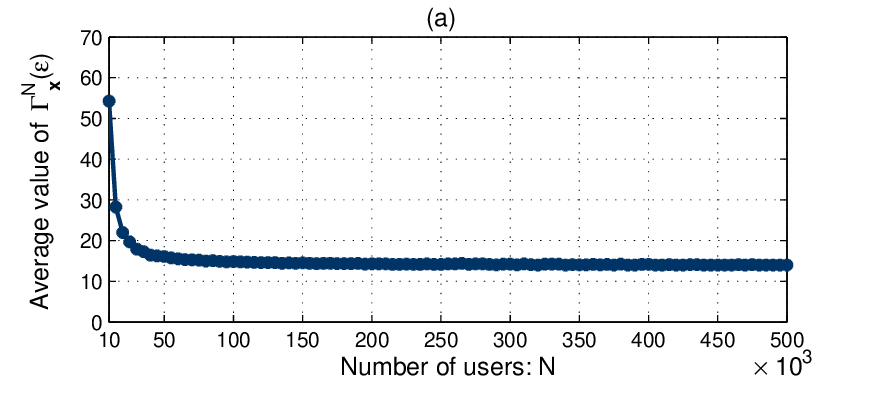}
\includegraphics[width=3.8in]{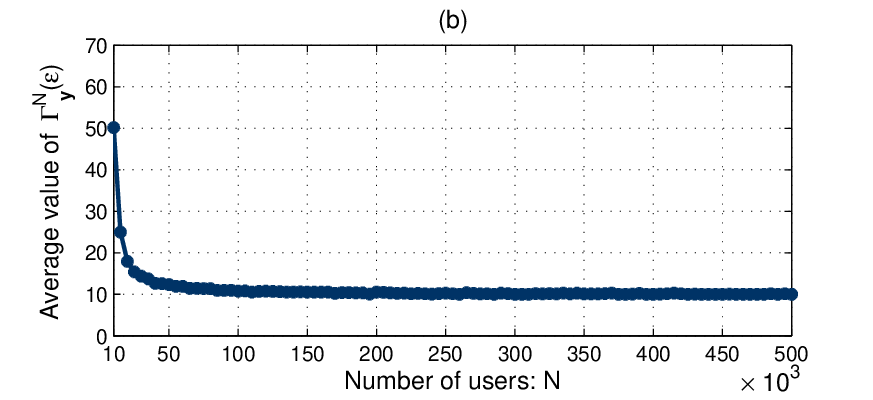}
\caption{Average time of hitting $\Omega_{\epsilon}(\vec{\bm \zeta}^{\alpha}_{\bm \gamma})$. (a) ${\bm Z}^N[0]=\bm x$; (b) $\bm Z^N[0]=\bm y$. }
\vspace{-8pt}
\label{fig:recur_time}
\end{figure}

\noindent \textbf{Remarks: }


1) We would like to emphasize that the global optimality result is
not a straight-forward extension of the local convergence result by
contrasting Proposition~\ref{prop:local_conv} and
Proposition~\ref{prop:asymp}. Note that in
Proposition~\ref{prop:local_conv}, the time limit is outside the
limit of the number of users $N$, where each convergence (with $N$)
is with respective to a \emph{fixed time duration}. However, the
order of limit is switched in the global optimality result of
Proposition~\ref{prop:asymp}, as it states the convergence with $N$
\emph{the infinite horizon} average throughput, which is much
stronger and hence is much more challenging to prove.

2) We would like to contrast Assumption $\Psi$ with Weber's
condition \cite{Weber}. For general RMBP problem, Weber's condition
leads to the same global asymptotic optimality result. While
confirming Weber's condition may be possible in very low-dimensional
problems, in our downlink scheduling problem, this requires one to
rule out the existence of both closed orbits and chaotic behavior of
a high-dimensional non-linear differential equation, which is
extremely difficult to check - even numerically. Assumption $\Psi$,
on the other hand, takes a much simpler form, as it is defined over
the actual stochastic system and is amenable to easy numerical
verification, as is performed in Section~\ref{sec:num:just}.


\begin{table*}
\vspace{3pt}
\begin{center}
\renewcommand{\tabcolsep}{.1cm}
\renewcommand{\arraystretch}{1}
\begin{tabular}{|c|c|c|c|c||c|c|c|c|c|}
\hline
\multicolumn{5}{|c||}{$\bm Z^N[0]=\bm x$} & \multicolumn{5}{|c|}{$\bm Z^N[0]=\bm y$}\\
\hline
$\alpha$ & $(p_1,r_1)$ & $(p_2,r_2)$ & $[\beta_1,\beta_2]$ & $E\big[\Gamma_{\bm x}^N
(\epsilon)\big]$ & $\alpha$ & $(p_1,r_1)$ & $(p_2,r_2)$ & $[\beta_1,\beta_2]$ & $E\big[\Gamma_{\bm y}^N
(\epsilon)\big]$\\
\hline
0.4360 & (0.2242,0.1379) & (0.6742,0.1376) & [0.6680,0.3320] & 24.8 & 0.1202 & (0.6598,0.0091) & (0.5881,0.1337)  &  [0.3534,0.6466]  & 50\\
\hline
0.0529 & (0.7209,0.2958) & (0.2393,0.0947) & [0.8772,0.1228] & 52.4 & 0.3857 & (0.5024,0.1382) & (0.1818,0.1442)  &  [0.8627,0.1373]  & 51\\
\hline
0.1368 & (0.6402,0.0611) & (0.9357,0.6544) & [0.9446,0.0554] & 20.8 & 0.8013 & (0.8335,0.2617) & (0.8046,0.1486)  &  [0.5621,0.4379]  & 9.8\\
\hline
0.6664 & (0.6016,0.0809) & (0.9163,0.2221) & [0.2571,0.7429] & 19.8 & 0.1410 & (0.5727,0.1403) & (0.0743,0.0418)  &  [0.4514,0.5486]  & 50\\
\hline
0.4558 & (0.8767,0.6747) & (0.8080,0.6483) & [0.6475,0.3525] & 5 & 0.6782 & (0.8871,0.0472) & (0.5157,0.0643) &  [0.2971,0.7029]  & 67.2\\
\hline
0.4606 & (0.9192,0.7814) & (0.2898,0.1686) & [0.9971	0.0029] & 15.8 & 0.0418 & (0.8311,0.0482) & (0.1699,0.0728)  &  [0.8828,0.1172]  & 60.6\\
\hline
0.1367 & (0.6401,0.0611) & (0.9357,0.6543) & [0.9446,0.0554] & 20.8 & 0.5858 & (0.4808,0.1552) & (0.8344,0.5340)  &  [0.4662,0.5338]  & 13 \\
\hline
0.6664 & (0.6016,0.0809) & (0.9163,0.2220) & [0.2571,0.7429] & 19.8 & 0.5271 & (0.7086,0.2569) & (0.8684,0.6064)  &  [0.7992,0.2008]  & 7.6\\
\hline
0.6018 & (0.2008,0.1861) & (0.2826,0.1992) & [0.7762,0.2238] & 3 & 0.8393 & (0.5426,0.1789) & (0.7747,0.4538)  &  [0.2453,0.7547]  & 5\\
\hline
0.1781 & (0.4421,0.0513) & (0.9150,0.4430) & [0.3696,0.6304] & 29 & 0.7498 & (0.5219,0.3849) & (0.6668,0.2956)  &  [0.9673,0.0327]  & 5.8\\
\hline
\end{tabular}
\end{center}
\caption{Evaluation of average time of hitting $\Omega_{\epsilon}(\vec{\bm \zeta}^{\alpha}_{\bm \gamma})$ under a wide range of parameters.}
\label{tab:tab4}
\vspace{-10pt}
\end{table*}

\section{Numerical Results}

\subsection{Verification and Interpretation of Assumption $\it \Psi$}
\label{sec:num:just}

We start by numerically verifying Assumption $\Psi$. We consider
the asymmetric scenario with two classes of channels with system
parameters $\gamma{=}[0.45, 0.55]$,
$\alpha{=}0.6$, with $p_1{=}0.9$, $r_1{=}0.45$, $p_2{=}0.8$, $r_2{=}0.3$.

We next examine the change of the average hitting time $\Gamma_{\bm x}^N(\epsilon)$, while maintaining $\alpha$ and $\bm \gamma$.

We let $\bm x, \bm y \in \mathcal{Z}$ be initial values of $\bm Z^N[0]$ that are selected to be two extreme points in the state space to exhibit the uniformity of $\Gamma^N_{\bm x}(\epsilon)$ to the initial state. Specifically, state $\bm x$ corresponds to the
case when all the users have just observed their channels to be in
OFF state, i.e., with belief value $b^k_{0,1}$, $k=1,2$. And $\bm y$
corresponds to the case when all users have no initial observation
of their channels state history, i.e., with belief value $b^k_s$,
$k=1,2$.

We examine the average value of hitting time $\Gamma_{\bm
x}^N(\epsilon)$ and $\Gamma^N_{\bm y}(\epsilon)$ with a very small
neighborhood $\epsilon{=}0.005$, when the number of users $N$ grows from $10{\times} 10^3$ to $500{\times}10^3$. As indicated in Fig.~\ref{fig:recur_time}, for both cases, the average time of hitting the $\epsilon$ neighborhood first decreases with $N$, and then \emph{converges} and stays almost the same as $N$ scales up. This is especially
intriguing. The rationale behind this phenomenon is as follows. Under Whittle's Index Policy, a total number of $\alpha N$ users are activated at each time slot. Therefore, for relatively small number of users, the amount of probabilistic belief state transitions, as well as the amount of system states in the neighborhood, increases with $N$, leading to a higher chance of hitting the desired neighborhood
$\Omega_{\epsilon}(\vec{\bm \zeta}^{\alpha}_{\bm \gamma})$ and smaller value of hitting time. However, the belief update of each user
contributes to the $1/N$ change of the system state $\bm Z^N[t]$, which
decreases with $N$. Therefore, as $N$ further increases, the \emph{total amount of transitions} of the system state ${\bm Z}^N[t]$ due to channel state feedback is roughly $\alpha N \cdot
1/N=\alpha$, which is invariant of $N$. Table~\ref{tab:tab4} illustrates the average value of hitting time $\Gamma_{\bm
x}^N(\epsilon)$ and $\Gamma^N_{\bm y}(\epsilon)$ under a variety of randomly generated system parameters when $1\%$ convergence is reached as $N$ scales. These result shows that the hitting time is bounded and hence of verifies Assumption $\Psi$.

\begin{figure}
\centering
\includegraphics[width=3.1in]{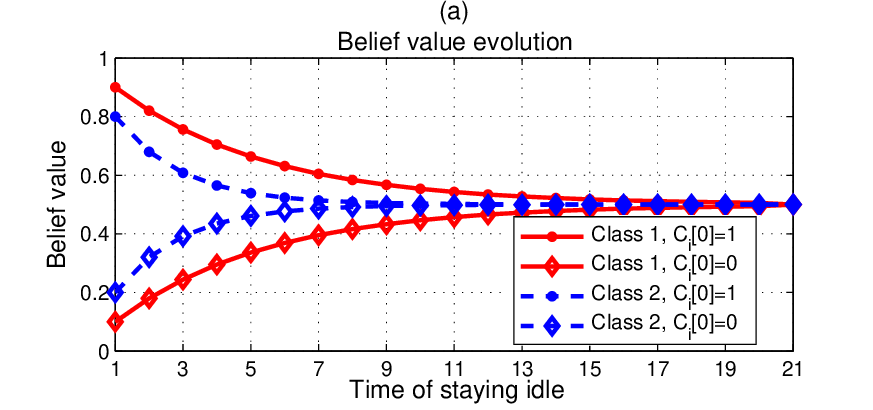}
\includegraphics[width=3.1in]{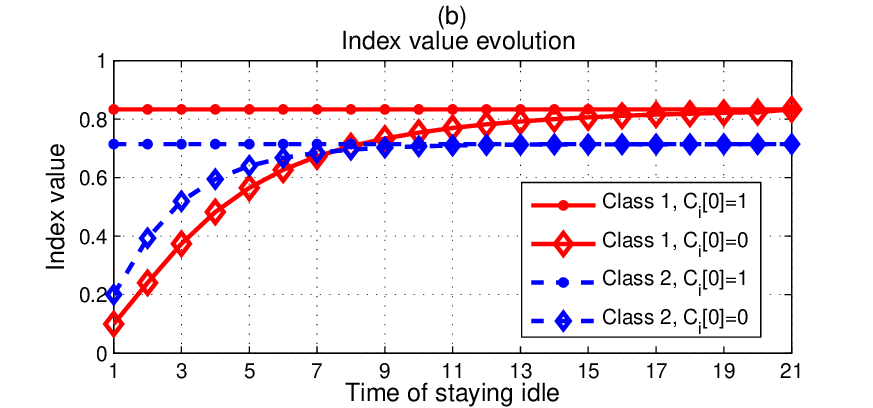}
\vspace{-3pt} \caption{The evolution of belief value and Whittle's
index value. (a)
Belief value evolution (b) Whittle's index value evolution. }
\vspace{-15pt}
\label{fig:beliefindex}
\end{figure}



\subsection{`Exploitation versus Exploration' Trade-off}
\label{sec:num:trade-off}

\vspace{-3pt}In this section, we demonstrate how the Whittle's index value captures
the `exploitation versus exploration' trade-off for our \emph{asymmetric
downlink scheduling problem}.

Consider two classes of ON/OFF fading channels with belief value
evolutions plotted in Fig.~\ref{fig:beliefindex}(a). Note that
both classes have the same stationary distribution $b^k_s=0.5$,
$k\in \{1,2\}$ of being at ON state, but channels in class $1$ has a higher
degree of time correlation, i.e., fades slower, than channels in
class $2$ since $p_1 > p_2$ and $r_1<r_2$. The corresponding Whittle
index values of the two classes of channels are depicted in
Fig.~\ref{fig:beliefindex}(b) as functions of the updated belief
value starting from different initial states.

To understand the nature of Whittle's index value, we first consider
the case when the channels in both classes are observed to be ON at
time $0$ and stay passive since then. As indicated in
Fig.~\ref{fig:beliefindex}(a) the class $1$ channel has a higher
belief value than the class $2$ channel, hence scheduling the
class $1$ channel gives a higher immediate throughput than
scheduling the class $2$ channel. Moreover, once a class $1$ channel
is scheduled, it is more likely to stay in ON state again,
bringing high future gains. Accordingly, the index values in
Fig.~\ref{fig:beliefindex}(b) when both state evolutions start from
ON states capture that it is more attractive to schedule the
class $1$ channel because of the advantage in both exploitation and
exploration.

On the other hand, when the scheduler has observed channels in both
classes to be OFF at time $0$, Fig.~\ref{fig:beliefindex}(a) shows
that the class $2$ channel has a higher belief value than the class
$1$ channel. However, although the Whittle's index value in
Fig.~\ref{fig:beliefindex}(b) of class $2$ channel is initially
smaller than that of class $1$ channel, after a certain amount of
delay (around $8$ slots in the figure) this order is switched, which
is interpreted as follows: initially, since the class $1$
channel has smaller belief value than that of the class $2$ channel,
it is more attractive to exploit the immediate gain brought by the
class $2$ channel. However, as the passive time grows, as indicated
in Fig.~\ref{fig:beliefindex}(a), the difference between immediate
gain of both classes diminishes. Then, it becomes more attractive to
explore the class $1$ channel because its longer memory can
bring higher future gains if it turns out to be in ON state.

This investigation reveals the intricate nature of Whittle's index
value in capturing the fundamental `exploration versus exploitation'
trade-off. In our scheduling problem with asymmetric channel
statistics, such a property of Whittle's Index Policy turns out to
be crucial in \emph{achieving asymptotically optimal performance}.

\begin{figure}
\centering
\includegraphics[width=3.7in]{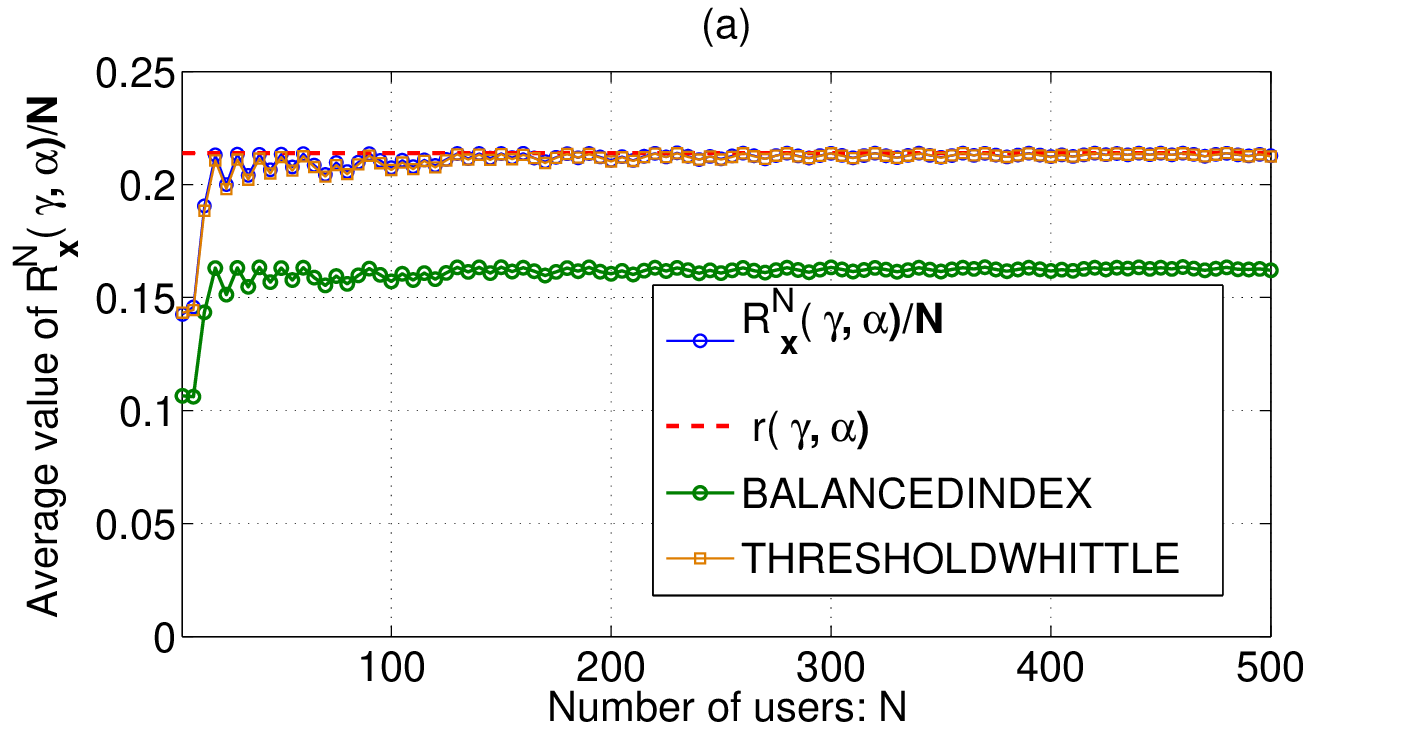}
\includegraphics[width=3.7in]{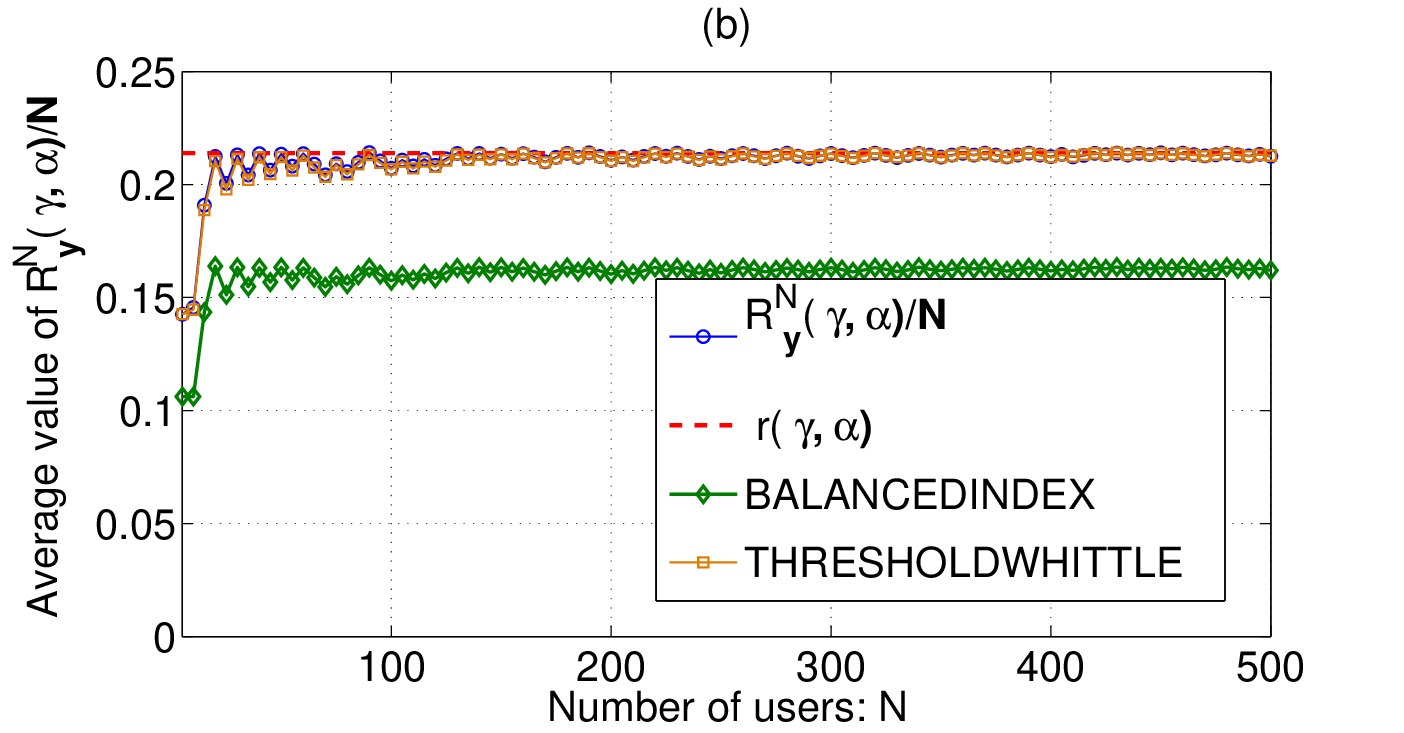}
\vspace{-3pt} \caption{Performance evaluation and comparison of per-user throughput of Whittle's Index Policy. (a) ${\bm Z}^N[0]=\bm x$; (b) $\bm Z^N[0]=\bm y$.}
\vspace{-15pt}
\label{fig:realistic}
\end{figure}

\subsection{Performance Evaluation and Comparison}

Note that our results focus on asymptotic regime when the number of users scales up. We next numerically evaluate the performance of the Whittle's Index Policy under finite number of users. We next consider a system where $\bm \gamma{=}[0.6, 0.4]$, $\alpha{=}0.3$, $(p_1,r_1){=}(0.75,0.2)$ and $(p_2,r_2){=}(0.8,0.3)$, and evaluate the value $R^{N}_{\bm x}(\bm \gamma, \alpha)/N$ when $N$ increases as multiples of $5$, i.e., $N=5m, m=1,2,\cdots$. Fig.~\ref{fig:realistic}(a) and (b) respectively correspond to the aforementioned extreme points. As observed in Fig.~\ref{fig:realistic}, the per-user throughput value $R^{N}_{\bm x}(\bm \gamma, \alpha)/N$ of Whittle's Index Policy quickly converges to the upper bound value $r(\bm \gamma, \alpha)$. This result indicates that, in realistic scenarios with finite $N$, the global convergence result in Proposition~\ref{prop:asymp} holds under moderate number of users (under $N=50$ as shown in Fig.~\ref{fig:realistic}). 

Fig.~\ref{fig:realistic} also plots the per-user throughput performance of the BALANCEDINDEX  policy, which is proposed in \cite{ApproxRMBP} and proved to achieve throughput half of the optimal throughput, i.e., 2-approximation performance. As observed in Fig.~\ref{fig:realistic}, the asymptotic per-user throughput performance of BALANCEDINDEX is strictly lower than the Whittle's Index Policy. This is because although BALANCEDINDEX policy guarantees 2-approximation to the optimal throughput performance, it does not provide strictly optimal per-user throughput performance in the asymptotic regime of large number of users, as compared with Whittle's Index Policy. Fig.~\ref{fig:realistic} also evaluates the performance of a slight modification Whittle's Index Policy, namely the THRESHOLD-WHITTLE policy, proposed in \cite{ApproxRMBP} by slightly adjusting the Whittles index value at belief values $p_i, i=1,2$. It can be observed from the figure that the per-user throughput performance of THRESHOLD-WHITTLE policy is very close to that of the Whittle's Index Policy, indicating that the modification of the Whittle's indices in THRESHOLD-WHITTLE policy does not bring significantly change the throughput performance for the plotted example. It was proven in \cite{ApproxRMBP} that the THRESHOLD-WHITTLE policy achieves at least half of the optimal  throughput. However, analytically proving the asymptotic optimality of THRESHOLD-WHITTLE policy remains an open question.

\subsection{Evaluation of Fairness among Users}

\begin{figure}
\centering
\includegraphics[width=3.2in]{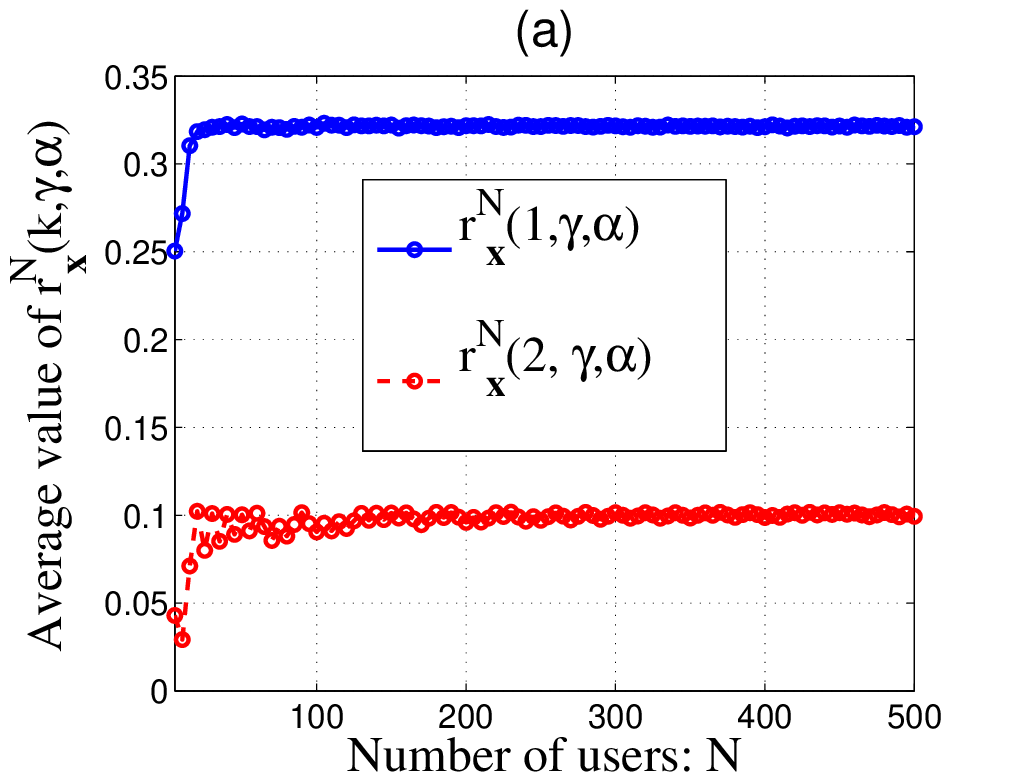}
\includegraphics[width=3.2in]{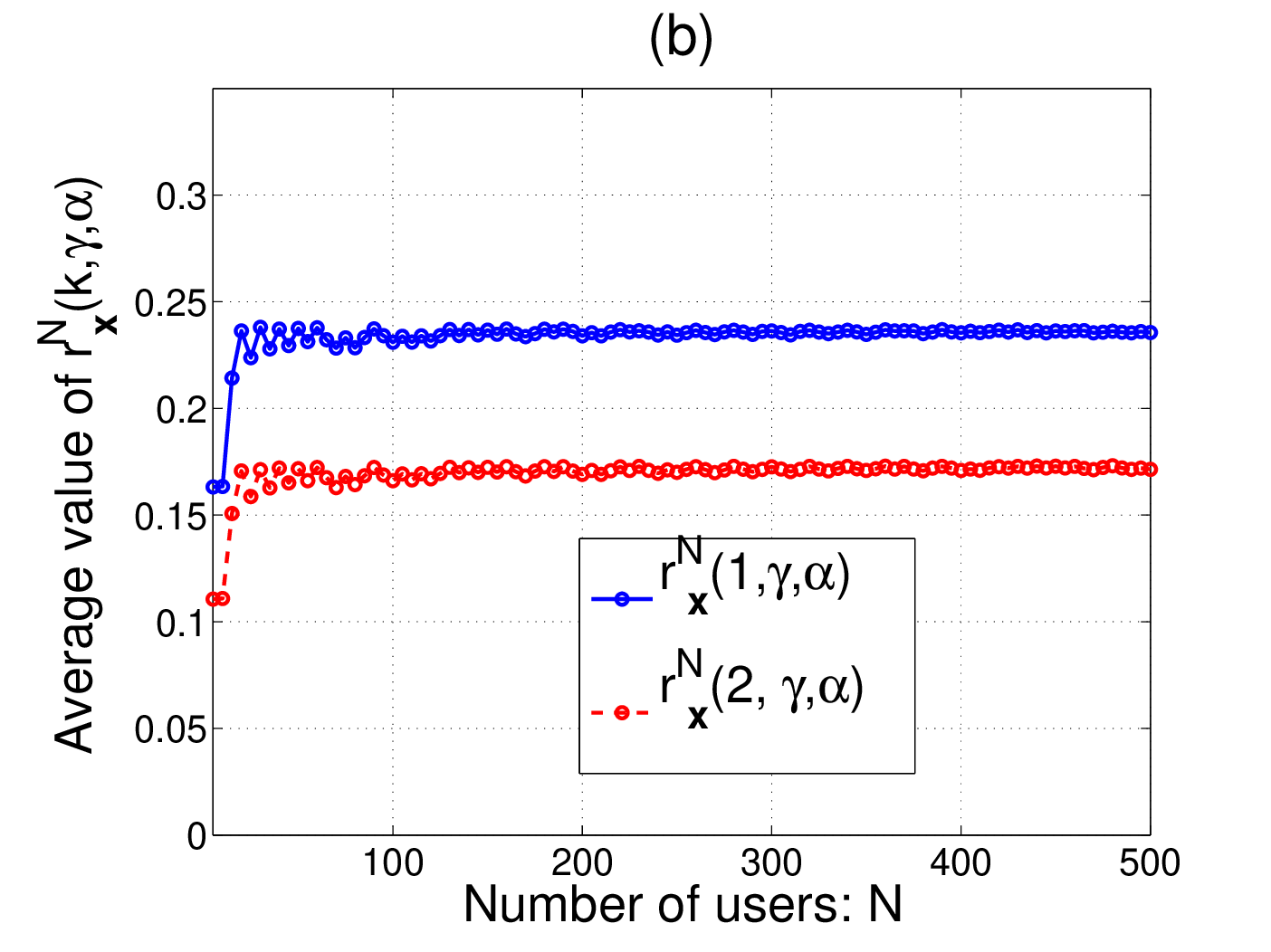}
\vspace{-3pt} \caption{Evaluation of $r^{N}_{\bm x}(k,\bm \gamma, \alpha)$ with $N$. (a) Whittle's Index Policy; (b) Policy $\Xi$.}
\vspace{-15pt}
\label{fig:Fairness}
\end{figure}

In this section, we evaluate the fairness performance of Whittle's Index Policy. We exam the throughput difference between the two types of users, under different set of Markov transition statistics. To facilitate better evaluation, we define the throughput $r^N_{\bm x}(k,\gamma,\alpha)$ to be the per-user throughput \emph{within each class $k$ of users}, i.e.,
\begin{align*}
&r^N_{\bm x}(k,\gamma,\alpha)\\
=&\frac{\lim_{T\rightarrow \infty}
\frac{1}{T}E \Big[\sum_{t=0}^{T-1} \sum_{i\in \mathcal{N}_k} \pi_i[t]
a_i^{ind}[t] \Big|{\bm Z}^{N}[0]\hspace{-2pt}=\hspace{-2pt}{\bm x} \Big]}{\gamma_k N},
\end{align*}
where $\mathcal{N}_k$ represents the set of users in class $k$. We consider the scenario where $(p_1,r_1)=(0.9,0.1)$ and $(p_2,r_2)=(0.6,0.4)$ with $\bm\gamma=[0.5,0.5], \alpha=0.3$. Therefore, the channels in class $1$ have a much higher degree of correlation than the channels in class $2$, i.e., it is more likely for the channels in class $1$ to stay in its previous-slot state than change to a different state compared with channels in class $2$. However, channels in both classes have the same steady state probability in state `1', i.e., $b_s^1=b_s^2=0.5$. Fig.~\ref{fig:Fairness} plots the per-user throughput within each class under Whittle's Index Policy. It can be observed that users in class $1$ achieves higher throughput than users in class $2$. The higher throughput gain of class $1$ is brought by the higher degree of temporal correlation and also the aforementioned `Exploitation versus Exploration' trade-off. Since the class-$1$ channels have higher degree of time-correlation, if a class-$1$ channel is previously observed in state $1$, the scheduler tends to continue to serve it for longer time to obtain high immediate gains. It is also more attractive to explore a channel in class $1$ because, as previously discussed, higher future gains can be obtained if it turns out to be in state `1'. Therefore, channels in class $1$ have higher overall throughput than channels in class $2$, resulting in the big gap in throughput between the two classes of users in Fig.~\ref{fig:Fairness}.

To facilitate better performance in terms of fairness, we evaluate the performance of the following heuristic policy $\Xi$ based on the Whittle's index values. In policy $\Xi$, instead of directly using Whittle's index values, the algorithm schedules the $\alpha N$ users with the largest
\begin{align*}
\frac{W_k(\pi_i[t])}{\overline{R}_i[t]},
\end{align*}
at slot $t$, where $\overline{R}_i[t]$ is user $i$'s achieved throughput up to slot $t$, i.e., $\overline{R}_i[t]=\sum_{\tau=1}^{t-1}\pi_i[\tau]\cdot a_i^{\Xi}[\tau]\big | \vec{\bm\pi}[0]$. Hence a user's priority for scheduling is determined by its Whittle's index value relative to its own actual achieved throughput. Therefore policy $\Xi$ mimics the proportional fair scheduling algorithms (e.g.,  \cite{Tse}) commonly used in communication networks. Fig.~\ref{fig:Fairness}(b) evaluates the performance of policy $\Xi$. As we can see, under the algorithm $\Xi$, the throughput gap between the two classes of channels is closer than Whittle's index policy, indicating improved fairness performance. Finally, we believe that combining Whittle's index and the frame-based scheduling \cite{Neely_utility} can lead to  low-complexity algorithms that optimally meet the fairness constraints among different users.



\section{Conclusion}
In this paper, we studied the problem of downlink scheduling over ON/OFF Markovian fading channels in the presence of channel heterogeneity. We consider the scenario where instantaneous channel state information is not perfectly known at the scheduler, but is acquired via a practical ARQ-styled feedback after each scheduled transmission. We analytically characterized the performance of Whittle's Index Policy for downlink scheduling, and proved its local and global asymptotic optimality properties as the number of users scales. Specifically, provided that the initial system state is within a certain region, we established the local optimality of Whittle's Index Policy by investigating the evolution of the system belief state with a fluid approximation. We then established the global asymptotic optimality of Whittle's Index Policy under a recurrence condition, which is suitable for numerical verification. Our results establish that Whittle's Index Policy, which is attractive due to its low-complexity operation, also processes strong asymptotic optimality properties for scheduling over heterogeneous Markovian fading channels. Future research directions includes design of scheduling algorithms that not only maximizes the sum throughput, but also provides fairness among heterogeneous users using Whittle's index.
\vspace{6pt}

\appendices

\section{Proof of Lemma~\ref{lemma:pos_rec}}
\label{appen:recur}

\begin{figure}
\centering
\includegraphics[width=3.3in]{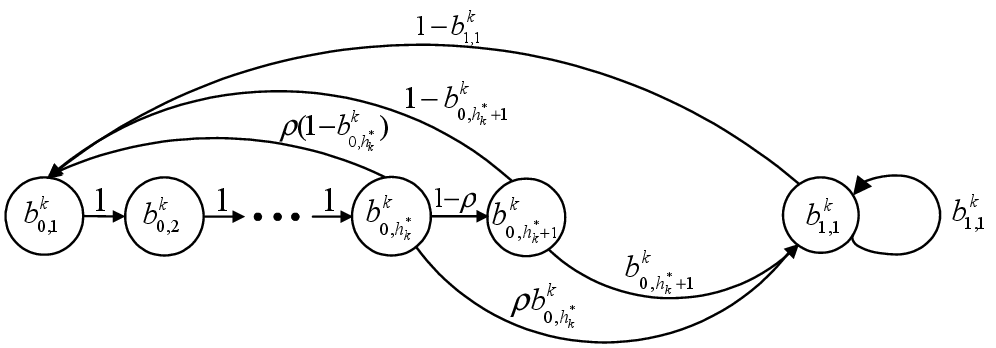}
\vspace{-4pt}
\caption{Belief value transition in steady state when $\omega^* = W_k(b^k_{0,h^*_k})$}
\vspace{-3pt}
\label{fig:belief_pos}
\end{figure}

(i) First consider the scenario where $\omega^* < W_k(b^k_s)$ and suppose $\omega^* = W_k(b^k_{0,h^*_k})$ for the belief state $b^k_{0,h^*_k}$. If the belief value of a channel is above $b^k_{0,h^*_k}$ at the beginning of a slot, the channel will be activated. According to the belief value evolution rule~(\ref{eq:evolve}), in the next slot its belief value will either be $p_k$ or $r_k$, depending on the underlying channel state revealed at the end of a slot. Clearly, the belief evolution in this case is positive recurrent within a finite state space, i.e., the belief state can only take the values $p_k, r_k, b^k_{0,2}, \cdots, b^k_{0,h^*_k+1}$. On the other hand, if the belief value is below $b^k_{0,h^*_k}$, the channel remains idle and will activate once its belief value exceeds $b^k_{0,h^*_k}$. Fig.~\ref{fig:belief_pos} illustrates the belief evolution in steady state under this scenario.
\vspace{4pt}

(ii) Consider the scenario where $\omega^* \hspace{1pt}{\geq}\hspace{1pt}W_k(b^k_s)$. In this case, a channel is activated if its index value is above $\omega^*$. After transmission, if the channel is observed to be in OFF state, its belief value will transit to $r_k$ and stays idle until its index value crosses $\omega^*$. Since $\omega^* \hspace{1pt}{\geq}\hspace{1pt}W_k(b^k_s)$, it is clear from the belief value evolution (see Fig.~\ref{fig:Qupdate}) that, starting from $r_k$, the belief value will always be smaller than $b^k_s$. Hence the channel will stay idle at all times. On the other hand, if the channel is observed to be in ON state after transmission, the belief value will transit to $p_k$ and the channel will keep on transmitting until the underlying channel turns out to be in OFF state. Since we assumed $p_k<1$, the channel will ultimately be in OFF state and its belief value will transit to $r_k$ and stays in idle mode ever since. Therefore eventually no channel in class $k$ will be scheduled and the belief values will keep transit toward, but never reach, the steady state belief value $b^k_{s}$.

\section{Proof of Lemma~\ref{lemma:parameter}}
\label{appen:peruser}

Consider two systems with different total number of users but identical $\alpha$ and $\bm \gamma$. Suppose the first system has $N_1$ total number of users while the second system has $N_2$ number of users. For the first system with $N_1$ total number of users, suppose the policy $\phi^*$, specified in Lemma~\ref{lemma:thres_relax}, is optimal for the relaxed-constraint problem. For each channel $i$ in class $k$, we let $A^k_{\phi^*}$ denote the expected fraction of time of activation, i.e.,
\begin{align}
\nonumber
A^k_{\phi^*}=\limsup_{T\rightarrow \infty} \frac{1}{T}E\Big[\sum_{t=0}^{T-1} a^{\phi^*}_i[t] \Big].
\end{align}

Then, according to Lemma~\ref{lemma:thres_relax}(ii), the expected number of activated users satisfies
\begin{align}
\nonumber\gamma_1 N_1 \cdot A^1_{\phi^*}+ \gamma_2 N_1 \cdot A^2_{\phi^*}=\alpha N_1.
\end{align}

Now apply the same policy $\phi^*$ when the total number of users is $N_2$. Since $\phi^*$ schedules each channel independently, $A^1_{\phi^*}$ and $A^2_{\phi^*}$ does not change in this scenario. Therefore, the expected number of activated users is expressed as
\begin{align}
&\gamma_1 N_2 \cdot A^1_{\phi^*}+ \gamma_2 N_2 \cdot A^2_{\phi^*}\nonumber\\
=&\frac{N_2}{N_1}\big[\gamma_1 N_1 \cdot A^1_{\phi^*}+ \gamma_2 N_1 \cdot A^2_{\phi^*}\big]=\alpha N_2,\nonumber
\end{align}
hence the complementary slackness condition (i.e., Lemma~\ref{lemma:thres_relax}(ii)) for the relaxed-constraint problem is also satisfied under $\phi^*$, when the total number of users is $N_2$. Hence the policy $\phi^*$ satisfies both Lemma~\ref{lemma:thres_relax}(i) and (ii) under the total number of users $N_2$, and is an optimal policy for that scenario.

Therefore, fixing system parameters $(\bm \gamma, \alpha)$, for different number $N$ of users, the policy $\phi^*$ is always optimal. Since the policy $\phi^*$ schedules each channel independently, we let $\upsilon_k(\bm \gamma, \alpha)$ denote the expected reward contributed by each channel in class $k$. Hence we have
\begin{align}
\nonumber
\upsilon^N(\bm \gamma, \alpha)=N \gamma_1 \upsilon_1(\bm \gamma, \alpha)+N \gamma_2 \upsilon_2(\bm \gamma, \alpha).
\end{align}

Therefore the per-user throughput is
\begin{align}
\nonumber
\frac{\upsilon^N(\bm \gamma, \alpha)}{N}=\gamma_1 \upsilon_1(\bm \gamma, \alpha)+\gamma_2 \upsilon_2(\bm \gamma, \alpha),
\end{align}
which is independent of $N$. Hence the lemma is proven.

\section{Proof of Lemma~\ref{lemma:zeta_invar}}
\label{appen:zeta_invar}

Given system parameters $(\bm \gamma, \alpha)$, we know from the proof of Lemma~\ref{lemma:parameter} that the form of the Optimal Relaxed Policy, denoted by $\phi^*$, does not change with the number $N$ of users. Since $\phi^*$ schedules each channel independently, we let vector $\bm \varepsilon^k=[ \varepsilon_{0,1}^{k}, \cdots,
\varepsilon_{0,\tau}^{k}, \varepsilon_{s}^{k}, \varepsilon_{1,\tau}^{k},
\cdots,\varepsilon_{1,1}^{k} ]$ denote the steady state distribution of the belief value of a user in class $k$ under $\phi^*$, with $\varepsilon^k_{s}+\sum_{c,h} \varepsilon^k_{c,h}=1$. Therefore,
\begin{align}
E[{\bm Z}^N(\infty)]&= \frac{1}{N}[ \gamma_1 N \bm \varepsilon^1, \gamma_2 N \bm \varepsilon^2 ]=[ \gamma_1 \bm \varepsilon^1, \gamma_2 \bm \varepsilon^2 ]. \nonumber
\end{align}

Since $\phi^*$ is independent of $N$, $\bm \varepsilon^k$ is independent of $N$ for $k=1,2$. Therefore  $E[{\bm Z}^N(\infty)]$ is independent of the user number $N$, which proves the lemma.

\section{Proof of Proposition~\ref{prop:local_conv}}
\label{appen:local}

\subsection{Notations}

We shall denote the $i^{th}$ element of $\bm Z^N[t]$ as $Z^N_i[t]$,  and let $\beta_i$ denote the corresponding belief value. The index value corresponding to $\beta_i$ is denoted as $w_i$. In this proof, since we are fixing the system parameters $(\bm \gamma,\alpha)$, we shall drop the suffixes $\alpha$ and $\bm \gamma$ to denote $\vec{\bm \zeta}^{\alpha}_{\bm \gamma}$ as $\vec{\bm \zeta}$.

For ease of exposition, in this proof we assume $W_2(b^2_{0,h^*_2-1})<W_1(b^1_{0,h^*_1})=\omega^*<W_2(b^2_{0,h^*_2})$. Hence, in the Optimal Relaxed Policy, channels in class $1$ are activated when their belief values are above $b^1_{0,h^*_1}$ and stay idle if their belief values are below $b^1_{0,h^*_1}$, and activates with probability $\rho^* {\in} (0,1)$ at $b^1_{0,h^*_1}$.  For channels in class $2$, they are activated when their belief values no smaller than $b^2_{0,h^*_2}$ and stay idle otherwise.

\subsection{Transition properties of the system state}

We first investigate the belief transition structure of the system state $\bm Z^N[t]$ under the Whittle's Index Policy. It is clear that $\bm Z^N[t]$ evolves as a Markov Chain. We define the \emph{expected drift} $\nabla \bm Z^N[t]$ associated with the transition of $\bm Z^N[t]$ as follows,

\begin{align}
\nabla \bm Z^N[t] =E\big[\bm Z^N[t+1]- \bm Z^N[t] \hspace{1pt}\big|\hspace{1pt} \bm Z^N[t] \big]. \label{eq:drift_def}
\end{align}

For a channel with belief value $\beta_{\hspace{-1pt}i}$, we let $q_{i,j}^0$ and $q_{i,j}^1$ be the probability that its belief state changes to state $\beta_{\hspace{-2pt}j}$ under the idle and transmission actions, respectively. For example, if $\beta_{\hspace{-1pt}i}$ corresponds to belief value $b^1_{0,l}$, then $q^0_{i,i+1}=1$ if the channel stays idle, otherwise $q^1_{i,1}=1-b^1_{0,l}$ and $q^1_{i,2\tau+1}=b^1_{0,l}$, which corresponds to the probability of observed channel being $0$ or $1$, respectively. Under the Whittle's Index Policy, we let $g_i(\bm z)$ be the fraction of users in belief value $\beta_i$ that are activated,
\begin{align}
\label{eq:ran_factor}
\hspace{-8pt}g_i(\bm z){=}\hspace{-3pt}\begin{cases}
\min \Big\{\Big[\frac{\alpha{-}\sum_{w_j>w_i} z_j> z_i}{z_i}\Big]^+, 1 \Big\}, \text{if $z_i{\neq}0$}\\
1, \hspace{0.2in}\text{if $z_i = 0$ and $\alpha{-}\sum_{w_j>w_i} z_j{>}0$}\\
0, \hspace{0.2in}\text{if $z_i = 0$ and $\alpha{-}\sum_{w_j>w_i} z_j{\leq}0$}
\end{cases}
\end{align}
where $[\cdot]=\max\{0,\cdot\}$.  We use $q_{i,j}(\bm z)$ to denote the probability that the belief value of a channel transit from $\beta_i$ to $\beta_j$ under system state $\bm z$. Then
\begin{align}
\label{eq:qz}
q_{ij}(\bm z)=g_i(\bm z) q_{ij}^1+\big(1-g_i(\bm z)\big) q_{ij}^0,
\end{align}
with
\begin{align}
\nonumber
q_{ij}^1=\begin{cases}
\beta_i&\text{if $i\leq2\tau+1, j=2\tau+1$}\\
1-\beta_i&\text{if $i\leq2\tau+1, j=1$}\\
\beta_i&\text{if $2\tau{+}2{\leq} i{\leq}2(2\tau{+}1), j{=}2(2\tau{+}1)$}\\
1-\beta_i&\text{if $2\tau{+}2{\leq}i{\leq}2(2\tau{+}1), j{=}2\tau{+}2$}\\
0&\text{otherwise.}
\end{cases}
\end{align}
\begin{align}
\nonumber
q_{ij}^0=\begin{cases}
1&\text{if $i{\leq}2\tau$ or $2\tau{+}2{\leq}i{\leq}3\tau{+}1$, and $j{=}i{+}1$}\\
1&\text{if $\tau+2\leq i\leq 2\tau+1, j=i-1$}\\
1&\text{if $3\tau+3\leq i<2(2\tau+1), j=i-1$}\\
1&\text{if $i=\tau+1$ or $3\tau+2$, and $j=i$}\\
0&\text{otherwise.}
\end{cases}
\end{align}

We shall let $\bm e_{ii}=\vec{0}$, and let $\bm e_{ij}, i\neq j$ be a vector that has $-1$ at the $i^{th}$ element, $+1$ at the $j^{th}$ element, and $0$ at all other elements. Hence if a user changes its belief state from $\beta_i$ to $\beta_j$, the corresponding change of the system state $\bm Z^N[t]$ is in the direction of $\bm e_{ij}$ with scale $1/N$. Therefore, $\nabla \bm Z^N[t]$ is a composition of expected changes in each direction $\bm e_{ij}$. Suppose $\bm Z^N[t]= \bm z$, since the expected amount of change of $\bm Z^N[t]$ in direction $\bm e_{ij}$ is $z_i[t] q_{ij}(\bm z[t])$, the expected drift $\nabla \bm Z^N[t]$ can then be written as,
\begin{align}
\nabla \bm Z^N[t] \Big|_{\bm Z^N[t]=\bm z} =\sum_{i,j} z_i q_{ij}(\bm z) \cdot \bm e_{ij} := Q(\bm z)\bm z, \label{eq:dir_comp}
\end{align}
where the $(i,j)^{th}$ element of matrix $Q(\bm z)$ is
\begin{align}
\label{eq:matrixQ}
Q_{ij}(\bm z)= \begin{cases}
-\sum_{j\neq i}q_{ij}(\bm z) &\text{for $i=j$},\\
q_{ji}(\bm z) &\text{for $i\neq j$}.
\end{cases}
\end{align}

Note that, although the system state $\bm z$ can only take values on a lattice that depends on N, the matrix function $Q_{ij}(\bm z)$ is defined over more general space $\mathcal{Z}$. Based on this, we proceed to define a fluid approximation model.

\subsection{Fluid Approximation Model}

We consider a fluid approximation model $\bm z[t]$, which is defined by the following difference equation
\begin{align}
\label{eq:fluid}
\bm z[t+1]-\bm z[t]= Q(\bm z[t])\bm z[t].
\end{align}

Note that the right-hand-side is completely determined by equation (\ref{eq:ran_factor})-(\ref{eq:matrixQ}), as a function of $\bm z[t]$ and is independent of $N$. We denote $\bm z[t]$ as the `fluid approximation model' because $\bm z[t]$ is no longer restricted to take values on the lattice as with the case of the original system state $\bm Z^N[t]$, and $\bm z[t]$ evolves in the direction of the \emph{expected change} of the system state \footnote{Note that by `fluid' we mean fluid in users/channels instead of fluid with respective to time.}. Recall that the set $\mathcal{Z}$ is defined in equation (\ref{eq:beta}), we proceed with the following lemma.

\begin{lemma}
\label{lemma:WitinZ}
If $\bm z[0]\in \mathcal{Z}$, then $\bm z[t]\in \mathcal{Z}$ for all $t \geq 0$.
\end{lemma}

\noindent \textbf{Proof:}
Since from (\ref{eq:dir_comp}) we have
\begin{align}
\bm z[t+1]-\bm z[t] \Big|_{\bm z[t]=\bm z} = Q(\bm z[t])\bm z=\sum_{i,j} z_i[t] q_{ij}(\bm z[t]) \cdot \bm e_{ij}. \nonumber
\end{align}

Note that the belief values of a channel can only evolve within the belief states of class of the channel, hence for class $1$,
\begin{align}
&\sum_{i=1}^{2\tau+1} z_i[t+1]-\sum_{i=1}^{2\tau+1} z_i[t]\nonumber\\
=& \vec{\bm 1}^T \cdot \hspace{-7pt} \sum_{1{\leq}i,j\leq{2\tau\hspace{-0.5pt} {+}\hspace{-1.5pt} 1}}  \hspace{-4pt} z_i[t] q_{ij}(\bm z[t]) \bm e_{ij} \nonumber \\
=&\hspace{-4pt} \sum_{1{\leq}i,j\leq{2\tau\hspace{-0.5pt} {+}\hspace{-1.5pt} 1}} z_i[t] q_{ij}(\bm z[t]) \cdot (1-1)\nonumber \\
=&\hspace{1pt}0. \nonumber
\end{align}
where $\vec{\bm 1}$ is a vector with $1$ in each element. Similar result holds for class $2$. Since $z[0] {\in} \mathcal{Z}$, we have
\begin{align}
\sum_{i=1}^{2\tau+1} z_i[t]\equiv \gamma_1, \ \sum_{i=2\tau+2}^{2(2\tau+1)} z_i[t]\equiv \gamma_2, \ \forall t\geq 0. \nonumber
\end{align}

Also equation (\ref{eq:dir_comp})-(\ref{eq:fluid}) indicates that $z_i[t]{\geq} 0$ for all $t{\geq} 0$ if $\bm z[0] \in \mathcal{Z}$. Therefore $\bm z[t] {\in} \mathcal{Z}$ for all $t\geq 0$, establishing the lemma. $\hfill \blacksquare$
\vspace{6pt}

\begin{lemma}
\label{lemma:alpha_achieve}
Given $(\alpha, \gamma)$, there exists a unique parameter pair $(\omega^*,\rho^*)$ for the optimal policy $\phi^*$.
\end{lemma}

\noindent \textbf{Proof:}
For a single channel $i$ in class $k$, consider the policy where the channel activates if its belief value $\pi_i>b^k$, stays idle when $\pi_i<b^k$, and activates with probability $\rho$ when $\pi_i=b^k$, for some belief value $b^k$. From the belief value evolution we can calculate the expected time of activion, denoted by $A^k(b^k, \rho)$,
\begin{align}
&A^k(b^k, \rho)\nonumber\\
\nonumber=&\begin{cases}
1-\frac{(1-p_k)(h-\rho)}{\rho b^k_{0,h}+(1-\rho)b^k_{0,h+1}+(1-p_k)(h+1-\rho)} &\text{if $b^k=b^k_{0,h}$,} \\
0 &\text{if $\pi \geq b^k_s$.}
\end{cases}
\end{align}

It is clear from its expression that, given $b^k$, $A^k(b^k, \rho)$ is continuous with $\rho$. Also we have $A^k(b^k_{0,h}, 0)=A^k(b^k_{0,h+1}, 1)$. In addition, some simple algebra reveals that, given $b^k_{0,h}$, $A^k(b^k_{0,h}, \rho)$ strictly increases with $\rho$. Therefore, since $A^k(b^k_{0,h}, 0)=A^k(b^k_{0,h+1}, 1)$, given $\rho$ $A^k(b^k, \rho)$ monotonically decreases with $b^k\in\mathcal{B}_k$.

Also, one can observe from the expression that, given $\rho$, $\lim_{h\rightarrow \infty} A^k(b^k_{0,h}, \rho)=0$ and $ A^k(b^k_{0,1}, 1)=1$. Hence by appropriately choosing $b^k$ and $\rho$, $A^k(b^k, \rho)$ can achieve any value within $[0,1]$.
\vspace{3pt}

Note that the index value $W_k(b^k)$ monotonically increases with $b^k \in \mathcal{B}_k$, $k=1,2$.  It follows from the above analysis that, as $\omega$ increases, under policy $\phi(\omega,1)$, the fraction of activation time for each user strictly decreases from $1$ to $0$. Therefore, there exists an unique $(\omega^*, \rho^*)$ pair, such that the policy $\phi(\omega^*, \rho^*)$ strictly satisfies activation constraint~(\ref{eq:relaxed}). $\hfill \blacksquare$

\begin{lemma}
\label{lemma:fixpt}
The vector $\vec{\bm \zeta}$ is the unique fixed point of the fluid approximation model, i.e., for all $\bm z \in \mathcal{Z}$,
$Q(\bm z)\bm z=0$ if and only if $\bm z=\vec{\bm \zeta}$.
\end{lemma}

\noindent \textbf{Proof:} The proof follows from a similar line of \cite{Weber}. Note that, under the Optimal Relaxed Policy, $\vec{\bm \zeta}=E\big[{\bm Z}^N(\infty)\big]$ and $\alpha$ fraction of channels are activated on average. Therefore, in the fluid approximation model, we have $\bm z[t+1]-\bm z[t] \big|_{\bm z[t]=\vec{\bm \zeta}} =0$, i.e., $Q(\vec{\bm \zeta})\vec{\bm \zeta}=0$.

Now suppose there exists another fixed point $\vec{\bm \zeta}_0\in \mathcal{Z}$ such that $\vec{\bm \zeta}_0 \neq \vec{\bm \zeta}$ and $Q(\vec{\bm \zeta}_0)\vec{\bm \zeta}_0=0$. Then $\vec{\bm \zeta}_0$ corresponds to the stationary distribution of the system state under another policy $\phi(\omega_0, \rho_0)$ with threshold parameter $\omega_0$ and randomization factor $\rho_0$. Furthermore, under $\phi(\omega_0, \rho_0)$, the expected fraction of activated channels equals to $\alpha$. However, this contradicts with Lemma~\ref{lemma:alpha_achieve}, which states that $(\omega^*, \rho^*)$ is the unique parameter pairs that strictly satisfies the average constraint of activation. Therefore, the fixed point $\vec{\bm \zeta}$ is unique.
 \hfill $\blacksquare$

\subsection{Convergence of the Fluid Limit Model}

Define the region $\mathcal{J}_{\omega^*} \subseteq \mathcal{Z}$ as the set of $\bm z \in \mathcal{Z}$ such that, under the Whittle's Index Policy defined in Section~\ref{sec:num:index}, the channel is activated if and only if its index value is no smaller than $\omega^*$, which is the threshold for the Optimal Relaxed Policy defined in Lemma~\ref{lemma:thres_relax}. This means that, at system state $z \in \mathcal{J}_{\omega^*}$, all channels with index value higher than $\omega^*$ are scheduled, and the channels with index value smaller than $\omega^*$ stay idle, while the channels at index value $\omega^*$ are scheduled with certain randomization. Specifically, $\mathcal{J_{\omega^*}}=\{\bm z {\in} \mathcal{Z}: \sum_{i: w_i > \omega^*} z_i < \alpha, \ \sum_{i: w_i \geq \omega^*}z_i  \geq \alpha.\}$.

The following lemma characterizes the linearity property of the fluid approximation model in $J_{\omega^*}$.
\vspace{3pt}

\begin{lemma}\label{lemma:piecewiseL}

\noindent(i) The vector  $\vec{\bm \zeta} \in {\mathcal{J}_{\omega^*}}$.

\noindent(ii) The fluid difference equation (\ref{eq:fluid}) is linear within the region $\mathcal{J}_{\omega^*}$, i.e., there exist matrix $\bm Q^*$ and vector $\bm a^*$ such that
\begin{align}
\bm z[t{+}1]{-}\bm z[t]{=}Q^* \cdot \bm z[t]{+}\bm a^*, \text{for all \ } \bm z[t] \in \mathcal{J}_{\omega^*}.\label{eq:Q_star}
\end{align}
\end{lemma}

\noindent \textbf{Proof:}
(i) The vector  $\vec{\bm \zeta} \in {\mathcal{J}_{\omega^*}}$ because, if $\bm z[t]=\vec{\bm \zeta}$, we have $\sum_{i: w_i \geq \omega^*} g_i(\bm z[t])z_i[t]=\alpha$, where $g_i(\bm z[t])z_i[t]\in [0,1]$ as defined in~(\ref{eq:ran_factor}).
\vspace{3pt}

(ii) Recall that, at the beginning of the section, we have assumed $\omega^*=W_1(b^1_{0,h^*_1})$ for the belief value $b^1_{0,h^*_1}$ of class-$1$ channel. The difference equation (\ref{eq:fluid}) becomes,
\begin{align}
\nonumber
&\bm z[t+1]-\bm z[t]\Big|_{\bm z[t]=\bm z} \nonumber\\
{=}& \sum_{i, j: i \neq h^*_1} z_i q_{ij}(\bm z) \cdot \bm e_{ij}+  z_{h^*_1} \sum_{j} q_{h^*_1 j}(\bm z) \cdot \bm e_{h^*_1 j} \nonumber \\
{=}& \sum_{i, j: i \neq h^*_1} z _i q_{ij}(\bm z ) \cdot \bm e_{ij}\nonumber\\
&\hspace{5pt}+ z_{h^*_1}  \sum_{j} \big[g_{h^*_1}(\bm z )q^1_{h^*_1 j}+[1{-}g_{h^*_1}(\bm z )]q^0_{h^*_1 j} \big] \cdot \bm e_{h^*_1 j}\nonumber \\
{=}&\sum_{i, j: i \neq h^*_1} z _i q_{ij}(\bm z ) \cdot \bm e_{ij}+  z_{h^*_1}  \sum_{j} q^0_{h^*_1 j}\cdot \bm e_{h^*_1 j}\nonumber\\
&\hspace{0.5in}+ g_{h^*_1}(\bm z) z_{h^*_1}  \sum_{j} \big[ q^1_{h^*_1 j}-q^0_{h^*_1 j} \big] \cdot \bm e_{h^*_1 j}. \label{eq:diff_linear}
\end{align}
where the second equality is from (\ref{eq:qz}).

Since the total fraction of users activated is $\alpha$, we have
\begin{align}
\label{eq:linear_alpha}
g_{h^*_1}(\bm z )z _{{h^*_1}} =\alpha-\sum_{w_i > \omega^*} z_i,
\end{align}

Substituting the expression (\ref{eq:linear_alpha}) back in (\ref{eq:diff_linear}), and noting that $q_{ij}(\bm z), i {\neq} h^*_1$ stays constant for $\bm z {\in} \mathcal{J}_{\omega^*}$ (since the threshold $\omega^*$ for activation does not change for $\bm z \in \mathcal{J}_{\omega^*}$), the linearity property holds. \hfill $\blacksquare$

\vspace{10pt}


From Lemma~\ref{lemma:WitinZ} we know that $\bm z[t] \in \mathcal{Z}$ for all $t \geq 0$, i.e.,
\begin{align}
\sum_{i=1}^{2\tau+1} z_i=\gamma_1, \quad \sum_{i=2\tau+2}^{2(2\tau+1)} z_i=\gamma_2 . \label{eq:beta2}
\end{align}

Taking note of Lemma~\ref{lemma:WitinZ}, instead of using a $2(2\tau+1)$ dimensional vector $\bm z$, it suffices to represent the system state by a $2 \cdot 2\tau$ dimension vector $\tilde{\bm z}$, i.e.,
\begin{align}
&\tilde{\bm z}=\big[z_1, \cdots, z_{h_1^*-1}, z_{h_1^*+1}, \cdots, z_{2\tau+h_2^*-1}, z_{2\tau+h_2^*+1},\nonumber\\
&\hspace{2.5in} \cdots, z_{2(2\tau+1)}]. \nonumber
\end{align}
in which elements $z_{h_1^*}$ and $z_{2\tau+h_2^*}$ are eliminated from $\bm z$. The transition of $ \tilde{\bm z}[t]$, when $\bm z[t] \in \mathcal{J}_{\omega^*}$, is obtained by substituting the relationship (\ref{eq:beta2}) in the difference equation (\ref{eq:diff_linear}) and eliminate the elements $z_{h_1^*}$ and $z_{2\tau+h_2^*}$, i.e.,
\begin{align}
\tilde{\bm z}[t+1]-\tilde{\bm z}[t]= U^* \cdot \tilde{\bm z}[t] + \bm b^*. \label{eq:Q_tilde},
\end{align}
where the matrix $U^*$ and vector $\bm b^*$ are obtained after the substitution.
The next key lemma captures the eigen structure of matrix $U^*$.

\begin{lemma}
\label{lemma:EigVal}
Each eigen value $\lambda$ of $U^*$ satisfies $\big|\lambda+1\big|<1$.
\end{lemma}

\begin{proof}
The proof is based on explicit study of matrix $U^*$ and is given in Appendix~\ref{appen:EigVal}.
\end{proof}
\vspace{4pt}

This lemma leads to the local convergence of $\bm z[t]$.

\begin{lemma}
\label{prof:fluid_conv}
There exists a positive constant $\sigma$ such that, if the initial state $\bm z[0]=\bm x$ of the fluid approximation model is within the $\sigma$ neighborhood $\Omega_{\sigma}(\vec{\bm \zeta})$ of $\vec{\bm \zeta}$, where $\Omega_{\sigma}(\vec{\bm \zeta}) \subseteq J_{\omega^*}$, then
\vspace{4pt}

(i) $\bm z[t] \in \mathcal{J}_{\omega^*}$ for all $t\geq0$;

(ii) $\bm z[t] \rightarrow \vec{\bm \zeta}$ as $t \rightarrow \infty$.
\end{lemma}

\noindent \textbf{Proof:} Similar to $\vec{\bm \zeta}$ that corresponds to ${\bm z}[t]$, we let vector ${\tilde{\bm \zeta}}$ represent the stationary expectation of vector $\tilde{\bm z}[t]$. Therefore, from Lemma~\ref{lemma:fixpt},
\begin{align}
U^* \cdot {\tilde{\bm \zeta}} + \bm b^*=0. \label{eq:Ustation}
\end{align}

Substituting (\ref{eq:Ustation}) in equation~(\ref{eq:Q_tilde}), we have
\begin{align}
\tilde{\bm z}[t]- \tilde{\bm \zeta}&=(U^*+I\big)^{t} (\bm z[0]- {\tilde{\bm \zeta}})\nonumber\\
&=(U^*+I\big)^{t} (\bm x- {\tilde{\bm \zeta}}). \label{eq:Ztilde}
\end{align}

Since we have assumed that $\rho^*\neq 1$, there exists a $\sigma_0$ neighborhood $\Omega_{\sigma_0}(\vec{\bm \zeta})$ with $\Omega_{\sigma_0}(\vec{\bm \zeta}) \subseteq J_{\omega^*}$. Correspondingly, there is a neighborhood of $\tilde{\bm \zeta}$ for which $\tilde{\bm z}[t]$ evolution is linear and is described by~(\ref{eq:Ztilde}). From Lemma~\ref{prof:fluid_conv}, each eigen value $\lambda$ of $(U^*+I)$ satisfies $\big|\lambda\big|<1$. According to the stability theory of linear systems \cite{Khalil}, $\tilde{\bm z}[t]$ converges to $\tilde{\bm \zeta}$ if the initial state is close enough to $\tilde{\bm \zeta}$.

Therefore, there exists a $\sigma<\sigma_0$ neighborhood of $\vec{\bm \zeta}$ for which if the initial state $\bm x \in \Omega_{\sigma}(\vec{\bm \zeta})$, $\bm z[t] \in  \mathcal{J}_{\omega^*}$ and $\bm z[t] \rightarrow \vec{\bm \zeta}$ as $t \rightarrow \infty$. $\hfill \blacksquare$

\subsection{Convergence of the system state}

The fluid approximation model provides a good estimate for the system state evolution when the number of users is large, captured in the following proposition, which can be viewed as a \emph{discrete-time version} of Kurtz theorem \cite{Kurtz} applied to our problem. The proof is given in Appendix~\ref{appen:discrete_kurtz}.

\begin{proposition}
\label{prop:discrete_kurtz}
There exists a neighborhood $\Omega_{\delta}(\vec{\bm \zeta})$ of $\vec{\bm \zeta}$ such that if $\bm Z^{N}[0]{=}\bm z[0]{=}\bm x \in \Omega_{\delta}(\vec{\bm \zeta})$, then for any $\mu>0$ and finite time horizon $T$ there exists positive constants $C_1$ and $C_2$ such that
\begin{align}
P_{\bm x}\Big( \sup_{0 \leq t < T} \big\| \bm Z^{N}[t]-\bm z[t] \big\| \geq \mu \Big)\leq C_1 \exp(-N C_2), \nonumber
\end{align}
where $\delta<\sigma$, and $P_{\bm x}$ denotes the probability conditioned on the initial state $\bm Z^{N}[0]=\bm x$. Furthermore, $C_1$ and $C_2$ are independent of $\bm x$ and $N$.
\end{proposition}

According to Proposition~\ref{prop:discrete_kurtz}, the system state $\bm Z^N[t]$ behaves very close to the fluid approximation model $\bm z[t]$ when the number of users $N$ is large. Since we have shown the convergence of $\bm z[t]$ to $\vec{\bm \zeta}$ within $\Omega_{\sigma}(\vec{\bm \zeta})$ in Lemma~\ref{prof:fluid_conv}, we are ready to establish the local convergence of the system state $\bm Z^N[t]$ to $\vec{\bm \zeta}$.

\begin{lemma}
\label{lemma:Kurtz_Index}
If $\bm Z^N[0]=\bm x \in \Omega_{\delta}(\vec{\bm \zeta})$, then for any $\mu>0$ there exists a time $T_0$ such that for each $T >T_0$, there exist positive constants $s_1$ and $s_2$ with,
\begin{align}
P_{\bm x}\Big( \sup_{T_0 \leq t < T} ||\bm Z^{N}[t]-\vec{\bm \zeta} || \geq \mu \Big)\leq s_1 \exp(-N s_2). \nonumber
\end{align}
\end{lemma}

\noindent \textbf{Proof:}
We let $0<\nu<\mu$. Noting that $\delta<\sigma$, from Lemma~\ref{prof:fluid_conv} we have, given $\bm z[0]=\bm x\in\Omega_{\delta}(\vec{\bm \zeta})$, there exists $T_0$ such that for all $t \geq T_0$.
\begin{align}
\nonumber
\big\|\bm z[t]- \vec{\bm \zeta}\big\| \leq \nu.
\end{align}

From Proposition~\ref{prop:discrete_kurtz} we know that there exist positive constants $s_1$ and $s_2$ such that,
\begin{align}
&P_{\bm x}\Big(\sup_{T_0 \leq t < T} \big\| \bm Z^{{N}}[t]{-}\vec{\bm \zeta} \big\| \geq \mu \Big)\nonumber\\
\leq& P_{\bm x}(\sup_{T_0 \leq t < T} \big\|\bm Z^{N}[t]{-}\bm z[t]\big\|{+}\big\|\bm z[t] {-} \vec{\bm \zeta}\big\|\geq \mu) \nonumber\\
\leq& P_{\bm x}(\sup_{T_0 \leq t < T} \big\|\bm Z^{N}[t]{-}\bm z[t]\big\|\geq \mu{-}\nu) \nonumber \\
\leq &  P_{\bm x}(\sup_{0 \leq t < T} \big\|\bm Z^{N}[t]{-}\bm z[t]\big\|\geq \mu{-}\nu) \nonumber \\
\leq &  s_1 \exp(-N s_2). \nonumber
\end{align}

Hence the lemma holds. $\hfill \blacksquare$
\vspace{3pt}

The previous lemma allows us to establish the local convergence result. Let $v: \mathcal{Z}\rightarrow \mathcal{R}$ be a mapping such that $v(\bm z)$ represents the per-user average throughput under system state $\bm z$. Therefore, $N v(\bm Z^N[t])$ is the immediate reward at time $t$ and we also have $r(\bm \gamma, \alpha)=v(\vec{\bm \zeta})$.

For $\ell>0$, we let $\mu>0$ be such that for any $\bm x\in \mathcal{Z}$, if $\|\bm x -\vec{\bm \zeta} \|<\mu$, then
\begin{align}
|v(\bm x)-v(\vec{\bm \zeta})|<\ell. \label{eq:ell_ineq}
\end{align}

Note that the per-user instantaneous throughput $v(\bm z)\leq 1$ and $T_0$ is defined in Lemma~\ref{lemma:Kurtz_Index}. Therefore,
\begin{align}
&\Big| \frac{R_{T}^{{N_m}}(\bm \gamma, \alpha, \bm x)}{N_m}-r(\bm \gamma, \alpha)\Big|\nonumber\\
=&\Big | \frac{1}{{N_m} T}E\big[\sum_{t=0}^{T-1} N_m v(Z^{N_m}[t])\big]-r(\bm \gamma, \alpha)\Big| \nonumber \\
=&\Big| \frac{1}{T}\sum_{t=0}^{T_0-1} E\big[v(Z^{N_m}[t])-v(\vec{\bm \zeta})\big]\nonumber\\
&\hspace{1.1in}+\frac{1}{T}\sum_{t=T_0}^{T-1}E \big[v(Z^{N_m}[t])-v(\vec{\bm \zeta})\big]\Big| \nonumber \\
\leq&  \Big| \frac{1}{T}\sum_{t=0}^{T_0-1} E\big[v(Z^{N_m}[t])-v(\vec{\bm \zeta})\big] \Big|\nonumber\\
&\hspace{1.1in}+\Big |  \frac{1}{T}\sum_{t=T_0}^{T-1}E \big[v(Z^{N_m}[t])-v(\vec{\bm \zeta})\big]\Big| \nonumber \\
\leq&  \frac{T_0}{T}+\frac{1}{T}\sum_{t=T_0}^{T-1}E\big[\big | v(Z^{N_m}[t])-v(\vec{\bm \zeta})\big| \big]. \label{eq:Tbound}
\end{align}

Letting $A_{N_m}$ be the event $\{\sup_{T_0 \leq t \leq T} ||\bm Z^{N_m}[t]-\vec{\bm \zeta} || \geq \mu\}$, we proceed to bound the second term in~(\ref{eq:Tbound}),
\begin{align}
&\frac{1}{T}\sum_{t=T_0}^{T-1}E\Big[ \big | v(Z^{N_m}[t])-v(\vec{\bm \zeta})\big| \Big ]\nonumber \\
=&P_{\bm x}(A_{N_m}) \frac{1}{T}\sum_{t=T_0}^{T-1}E\Big[ \big | v(Z^{N_m}[t])-v(\vec{\bm \zeta})\big| \Big | A_{N_m} \Big]+\nonumber\\
&\hspace{0.1in} \big(1-P_{\bm x}(A_{N_m})\big) \frac{1}{T}\sum_{t=T_0}^{T-1}E\Big[ \big | v(Z^{N_m}[t])-v(\vec{\bm \zeta})\big| \Big | \bar{A}_{N_m} \Big] \nonumber \\
\leq&  P_{\bm x}(A_{N_m})+ (1-P_{\bm x}(A_{N_m}))\ell \nonumber \\
=& P_{\bm x}(A_{N_m})(1-\ell) +\ell .\nonumber
\end{align}
where the inequality if from the fact $v(\bm z)\leq 1$ and the relation~(\ref{eq:ell_ineq}).

According to Lemma~\ref{lemma:Kurtz_Index}, when $\bm x \in \Omega_{\delta}(\vec{\bm \zeta})$, we have $\lim_{m \rightarrow \infty} P_{\bm x}(A_{N_m})=0$, therefore,
\begin{align}
\lim_{m \rightarrow \infty} \Big| \frac{R_{T}^{{N_m}}(\bm \gamma, \alpha, \bm x)}{{N_m}}-r(\bm \gamma, \alpha)\Big|& \leq  \frac{T_0}{T} + \ell. \nonumber
\end{align}

Since $\ell$ can be arbitrarily small, we have
\begin{align}
\lim_{m \rightarrow \infty}  | \frac{R_{T}^{{N_m}}(\bm \gamma, \alpha, \bm x)}{{N_m}}-r(\bm \gamma, \alpha)|\leq \frac{T_0}{T}. \nonumber
\end{align}

Hence, taking limit with $T$ in both sides,
\begin{align}
\lim_{T \rightarrow \infty} \lim_{m \rightarrow \infty}  \frac{R_{T}^{N_m}(\bm \gamma, \alpha, \bm x)}{N_m}\hspace{1pt}{=}\hspace{1pt}r(\bm \gamma, \alpha) \nonumber.
\end{align}

We have thus proved Proposition~\ref{prop:local_conv}.

\section{Proof of Lemma~\ref{lemma:recur}}
\label{appen:recur}

(i) Here we prove the Markov chain has one unique class by stating that, starting from any state, there exists a possibility to reach a particular state, and hence there is only one class of recurrent state. Without loss of generality, we assume $W_1(b^1_{1,1})\geq W_2(b^2_{1,1})$.

Case (1). Suppose $\alpha\leq \gamma_1$.  Starting from any initial state $\bm Z^N[0]$, the following transition can occur: whenever the channels in class $1$ are activated, their states are observed to be in ON state, and whenever channels in class $2$ are activated, they are revealed to be in OFF state. Then after a long enough time duration $t_1$, $\alpha$ fraction of channels, which are in class $1$, will be in belief value $p_1$, and other channels will have stationary belief value $\pi_s$. Hence the system state will be $\bm Z^N[t_1]=[\bm Z^{1,N}[t_1], \bm Z^{2,N}[t_1]]$ (defined in Section \ref{sec:local:Z}) with $Z^{1,N}_{1,1}[t_1]=\alpha$, $Z^{1,N}_{s}[t_1]=\gamma_1-\alpha$, $Z^{2,N}_{s}[t_1]=\gamma_2$, and with $0$ in all other positions.

Case (2). Suppose $\alpha>\gamma_1$. Starting from any initial state $\bm Z^N[0]$, consider the following transition path. Within the first period of time slots, $0\leq t \leq t_0$, whenever users in class $1$ are activated, they turn out to be in state $1$, and whenever users in class $2$ are activated, they turn out to be in state $0$. Then if $t_0$ is long enough, $\bm Z^{1,N}[t_0]$ is such that $Z^{1,N}_{1,1}[t_0]=\gamma_1$, with zero in all other elements. In the second period, $t_0\leq t \leq t_1$, whenever users in class $1$ are activated, it will remain in state $1$, and whenever users in class $2$ are activated, it turns out to be in state $1$ as well. Then after long enough time until $t_1$,   $\bm Z^N[t_1]=[\bm Z^{1,N}[t_1], \bm Z^{2,N}[t_1]]$ with $Z^{1,N}_{1,1}[t_1]=\gamma_1$, $Z^{2,N}_{1,1}[t_1]=\alpha-\gamma_1$,  and $Z^{2,N}_{s}[t_1]=1-\alpha=\gamma_2-Z^{2,N}_{1,1}[t_1]$, with zero in all other elements.

Since the state space of the Markov Chain $\bm Z^N[t]$ is finite, there is at least one recurrent class. As we have seen in the above cases that, starting from all states, $\bm Z^N[t]$ can reach a particular state. Therefore there can only be one recurrent state. We shall henceforth denote this particular state as $\bm Z^N_p$. It is also clear from the proof that the Markov chain is aperiodic because of the possible self-transition in state $\bm Z^N_p$.

(ii) Similar to the proof of Proposition~\ref{prop:local_conv}, in this part, we drop the suffix $\alpha$ and $\bm \gamma$ in the notation $\vec{\bm \zeta}^{\alpha}_{\bm \gamma}$, and we assume, with no loss of generality, $W_2(b^2_{0,h^*_2-1})<W_1(b^1_{0,h^*_1})=\omega^*<W_2(b^2_{0,h^*_2})$. Recall that from the expression~(\ref{eq:indices}) of Whittle's index value that $W_k(\pi)=W_k(b^k_s)$ for $\pi \in \mathcal{B}_k$, $\pi\geq b^k_s$, $k=1,2$. We first characterize the structure of $\vec{\bm \zeta}$. From the description in Lemma~\ref{lemma:pos_rec} we know that the non-zero elements of $\vec{\bm \zeta}$ are
\begin{align}
\zeta^1_0:=&\zeta^1_{0,1}=\zeta^1_{0,2}=\cdots=\zeta^1_{0,h_1^*}, \zeta^1_{0,h_1^*{+}1}=(1- \rho^*) \zeta^1_{0,h_1^*}\nonumber\\
\zeta^1_{1,1}=&1-\sum_{h=1}^{h_1^*+1} \zeta^1_{0,h}, \nonumber \\
\zeta^2_0:=&\zeta^2_{0,1}=\zeta^2_{0,2}=\cdots=\zeta^2_{0,h_2^*-1} = \zeta^2_{0,h_2^*}\nonumber\\
\zeta^2_{1,1}=&1-\sum_{h=1}^{h_2^*} \zeta^2_{0,h}. \nonumber
\end{align}

We shall proceed to construct a path from the state $\bm Z^N_p$ to an arbitrary neighborhood of  $\vec{\bm \zeta}$. For ease of exposition, in the proof we no longer consider the channels as unsplittable entities. Instead, the transition in the each stages (in the following proof) deals with belief state evolution of certain \emph{fraction} of users. As we shall see, under this assumption, we can construct a transition path of $\bm Z^N[t]$ under the Whittle's Index Policy, that transits from $\bm Z^N_p$ to the \emph{exact} value $\vec{\bm \zeta}$. Although the identified path may not be feasible in reality for small value of $N$, but as the number of users $N$ increases, we can find a transition path, which operates each user as unsplittable entities, that is arbitrarily close to this identified path, and thus can ultimately get arbitrarily close to any neighborhood of $\vec{\bm \zeta}$.

Note that when $\bm Z^N[t_1]=\bm Z^N_p$, $\bm Z^N[t_1]=\big[\bm Z^{1,N}[t_1], \bm Z^{2,N}[t_1]\big]$, where
\begin{align}
Z^{1,N}_{1,1}[t_1]{+} Z^{1,N}_{s}[t_1]=\gamma_1, \hspace{6pt} Z^{2,N}_{1,1}[t_1]{+}Z^{2,N}_{s}[t_1]=\gamma_2. \nonumber
\end{align}

In the following construction we shall assume that belief values are updated at the end of each slot when the actual channel states are revealed.
\vspace{6pt}

\noindent \textbf{Case (1). }Suppose $h^*_1 \geq h^*_2$ and $W_1(b^1_s)\geq W_2(b^2_s)$. We shall denote $h_1'=\max\{l: W_1(b^1_{0,l}) \leq  W_2(b^2_s)\}$. In this case,
the path is constructed with the stages below, starting from state $\bm Z^N[t_1]=\bm Z^N_p$.
\vspace{7pt}

\noindent\textbf{Stage 1.1.} In the first slot, among the $\alpha$ fraction activated channels, $\alpha-\zeta^1_{0,h_1^*+1}$ amount remains in ON state, and $\zeta^1_{0,h_1^*+1}$ amount turn out in OFF state and are in class $1$. Hence the end of this slot, $\bm Z^N=[\bm Z^{1,N}, \bm Z^{2,N}]$ has the following non-zero elements
\begin{align}
&Z^{1,N}_{0,1}=\zeta^1_{0,h_1^*+1}, \quad Z^{1,N}_{1,1}+ Z^{1,N}_{s}=\gamma_1-\zeta^1_{0,h_1^*+1}\nonumber\\
&Z^{2,N}_{1,1}+Z^{2,N}_{s}=\gamma_2. \nonumber
\end{align}

\noindent\textbf{Stage 1.2.} In each of the next $h_1^*$ slots, $\alpha-\zeta^1_0$ amount in the activated channels turn out in ON state, and $\zeta^1_0$ amount of them turn out to be in OFF state and are in class $1$. So at the end of the last slot of this stage, the non-zero elements of the system state $\bm Z^N=[\bm Z^{1,N}, \bm Z^{2,N}]$ satisfies
\begin{align}
&Z^{1,N}_{0,1}=Z^{1,N}_{0,2}=\cdots=Z^{1,N}_{0,h_1^*} = \zeta^1_0, \hspace{4pt} Z^{1,N}_{0,h_1^*{+}1}=\zeta^1_{0,h_1^*{+}1}\nonumber\\
&Z^{1,N}_{1,1}=\zeta^1_{1,1}, Z^{2,N}_{1,1}+Z^{2,N}_s=\gamma_2. \nonumber
\end{align}

\noindent\textbf{Stage 2.} In the next few slots, all activated channels turn out to be in state $1$. This stage goes on for $h_1'-h_1^*$ slots, until those channels that reach belief state $b^1_{0,1}$ at the end of stage 1.1 are in belief state $b^1_{0,h_1'+1}$. Then by the end of the last slot of this stage, the non-zero elements of the system state $\bm Z^N$ satisfies
\begin{align}
&Z^{1,N}_{0,h_1'-h_1^*+1} =\cdots= Z^{1,N}_{0,h_1'}=\zeta^1_0, \hspace{4pt} Z^{1,N}_{0,h_1'+1} =\zeta^1_{0,h_1^*+1}\nonumber\\
&Z^{1,N}_{1,1}=\zeta^1_{1,1}, Z^{2,N}_{1,1}+Z^{2,N}_s=\gamma_2. \nonumber
\end{align}

\noindent\textbf{Stage 3.} In each of the following slots, among all channel activated, only those in belief state $b^1_{0,h_1'{+}1}$  turn out to be in OFF state. This stage goes on until those channels that transit to belief state $b^1_{0,h_1'}$ in stage 2 reaches belief state $b^1_{0, h_1^*-h_2^*+1}$. Hence by the end of the final slot of this stage,
\begin{align}
&Z^{1,N}_{0,1}=\cdots=Z^{1,N}_{0,h_1^*-h_2^*} = \zeta^1_0, Z^{1,N}_{0,h_1^*-h_2^*+1}=\zeta^1_{0,h_1^*{+}1}\nonumber\\
&Z^{1,N}_{0,h_1'-h_2^*+2}=\cdots=Z^{1,N}_{0,h_1'+1} = \zeta^1_0, Z^{2,N}_{1,1}+Z^{2,N}_s=\gamma_2. \nonumber
\end{align}

\noindent\textbf{Stage 4.} In each of the next $h_2^*$ slots, among all users activated, those in belief state $b^1_{0,h_1'{+}1}$  turn out to be in OFF state, and $\zeta^2_0$ amount of activated channels in class $2$ turn out in OFF state.
Then by the end of the final slot in this stage, the system state will be $\bm Z^N=\vec{\bm \zeta}$, i.e.,
\begin{align}
&Z^{1,N}_{0,1}=Z^{1,N}_{0,2}=\cdots=Z^{1,N}_{0,h_1^*}=\zeta^1_0\nonumber\\ &Z^{1,N}_{0,h_1^*{+}1}= \zeta^1_{0,h_1^*{+}1}, \ Z^{1,N}_{1,1}= \zeta^1_{1,1} \nonumber \\
&Z^{2,N}_{0,1}=Z^{2,N}_{0,2}=\cdots=Z^{2,N}_{0,h_2^*-1} = Z^{2,N}_{0,h_2^*}{=}\zeta^2_0, Z^{2,N}_{1,1}{=}\zeta^2_{1,1}. \nonumber
\end{align}
\vspace{2pt}

\noindent \textbf{Case (2).} Suppose $W_1(b^1_s)\geq W_2(b^2_s)$ and $h^*_1 \leq h^*_2$. We shall let $h_1'=\max\{l: W_1(b^1_{0,l}) \leq  W_2(b^2_s)\}$ and $d=\lfloor h^*_2 / (h_1'+1) \rfloor$. Starting from state $\bm Z^N[t_1]=\bm Z^N_p$, the path is constructed with the stages below, where stage 1.1 and 1.2 are the same with the previous case.
\hspace{10pt}

\noindent\textbf{Stage 1.1.} In the first slot, among the $\alpha$ fraction of activated channels, only $\zeta^1_{0,h_1^*+1}$ amount turn out in OFF state and they are in class $1$. Therefore at the end of this slot, $\bm Z^N=[\bm Z^{1,N}, \bm Z^{2,N}]$ with non-zero elements being
\begin{align}
&Z^{1,N}_{0,1}=\zeta^1_{0,h_1^*+1}, \quad Z^{1,N}_{1,1}+ Z^{1,N}_{s}=\gamma_1-\zeta^1_{0,h_1^*+1}\nonumber\\
&Z^{2,N}_{1,1}+Z^{2,N}_{s}=\gamma_2. \nonumber
\end{align}

\noindent\textbf{Stage 1.2.} In each of the next $h_1^*$ slots, $\alpha-\zeta^1_0$ amount of activated channels are in state `1', and $\zeta^1_0$ amount are in OFF state and are in class $1$. Hence at the end of the last slot of this stage, the non-zero elements of $\bm Z^N=[\bm Z^{1,N}, \bm Z^{2,N}]$ satisfies
\begin{align}
&Z^{1,N}_{0,1}=Z^{1,N}_{0,2}=\cdots=Z^{1,N}_{0,h_1^*} = \zeta^1_0, \hspace{4pt} Z^{1,N}_{0,h_1^*{+}1}=\zeta^1_{0,h_1^*{+}1}\nonumber\\
&Z^{1,N}_{1,1}=\zeta^1_{1,1}, \hspace{4pt} Z^{2,N}_{1,1}+Z^{2,N}_s=\gamma_2. \nonumber
\end{align}

Letting $t_2$ be the slot right after stage 1.2, the path proceeds as follows.
\vspace{7pt}

\noindent\textbf{Stage 2.}

\noindent(1) From slot $t_2$ to slot $t_2+h_1'-h_1^*-1$, all activated channels in class $1$ turn out to be in state $1$. Hence at the end of slot $t_2+h_1'-h_1^*-1$, the channels that reach belief state $b^1_{0,h_1^*{+}1}$ at the end of stage 1.2 are in belief state $b^1_{0,h_1'+1}$. Next, from slot $t_2+h_1'-h_1^*$ to slot $t_2+(d+1)(h_1'+1)-1$, among the activated channels in class $1$, only those in belief state $b^1_{0,h_1'{+}1}$  turn out to be in OFF state. Therefore, at the end of slot $t_2+(d+1)(h_1'+1)-1$, the system state vector $\bm Z^{1,N}$ that correspond to class-$1$ channels is
\begin{align}
&Z^{1,N}_{0,1}=Z^{1,N}_{0,2}=\cdots=Z^{1,N}_{0,h_1^*}=\zeta^1_0\nonumber\\
&Z^{1,N}_{0,h_1^*{+}1}= \zeta^1_{0,h_1^*{+}1}, \quad Z^{1,N}_{1,1}= \zeta^1_{1,1}. \nonumber
\end{align}

\noindent(2) In the meanwhile, from slot $t_2+(d+1)(h_1'+1)-h_2^*-1$ to slot $t_2+(d+1)(h_1'+1)-1$, among the activated channels in class $2$, $\zeta^2_0$ amount turn out to be in OFF state. Hence by the end of slot $t_2+(d+1)(h_1'+1)-1$, the vector $\bm Z^{1,N}$ that correspond to class-$2$ channels is
\begin{align}
Z^{2,N}_{0,1}{=}Z^{2,N}_{0,2}{=}\cdots{=}Z^{2,N}_{0,h_2^*-1}{=} Z^{2,N}_{0,h_2^*}{=}\zeta^2_0, Z^{2,N}_{1,1}{=} \zeta^2_{1,1}. \nonumber
\end{align}

Therefore, at the end of slot $t_2+(d+1)(h_1'+1)-1$,  $\bm Z^N=\vec{\bm \zeta}$.
\vspace{10pt}

\section{Proof of Lemma~\ref{lemma:steady_dist}}
\label{appen:Invar_meas}

The proof is a discrete-time version of the proof of Theorem 6.89 from \cite{Weiss_LD}. We first present a lemma which is an extension of Lemma~\ref{lemma:Kurtz_Index}.
\begin{lemma}
\label{lemma:Kurtz_Index3}
There is a neighborhood $\Omega_{\vartheta}(\vec{\bm \zeta}^{\alpha}_{\gamma})$ of $\vec{\bm \zeta}^{\alpha}_{\gamma}$, with $\vartheta<\delta$, for which if $\bm Z^N[0]=\bm x \in \Omega_{\vartheta}(\vec{\bm \zeta}^{\alpha}_{\gamma})$, then for any $\mu>0$ and time $T$, there exist positive constants $\rho_1$ and $\rho_2$ with,
\begin{align}
P_{\bm x}\Big( \sup_{0 \leq t < T} \|\bm Z^{N}[t]-\vec{\bm \zeta}^{\alpha}_{\gamma} \| \geq \mu \Big)\leq \rho_1 \exp(-N \rho_2) \nonumber
\end{align}
where $\rho_1$ and $\rho_2$ are independent of $\bm x$ and $N$.
\end{lemma}

\begin{proof}
Note that we have established, in Lemma~\ref{prof:fluid_conv}, the local convergence of the fluid approximation model $\bm z[t]$ in a neighborhood $\Omega_{\sigma}(\vec{\bm \zeta}^{\alpha}_{\gamma})$. We let $\nu < \mu$ and let $\vartheta<\delta$ (recall that $\delta$ is defined in proposition~\ref{prop:discrete_kurtz} with $\delta<\sigma$) be such that if $\bm z[0] \in \Omega_{\vartheta}(\vec{\bm \zeta}^{\alpha}_{\gamma})$, then
\begin{align}
\nonumber
\|\bm z[t]- \vec{\bm \zeta}^{\alpha}_{\gamma}\| \leq \nu, \quad \forall t\geq 0.
\end{align}

From Proposition~\ref{prop:discrete_kurtz}, there exist positive constants $\rho_1$ and $\rho_2$ with,
\begin{align}
&P_{\bm x}\Big(\sup_{0 \leq t < T} \big\| \bm Z^{{N}}[t]{-}\vec{\bm \zeta} \big\| \geq \mu \Big)\nonumber\\
\leq& P_{\bm x}(\sup_{0 \leq t < T} \big\|\bm Z^{N}[t]{-}\bm z[t]\big\|{+}\big\|\bm z[t] {-} \vec{\bm \zeta}\big\|\geq \mu) \nonumber\\
\leq& P_{\bm x}(\sup_{0 \leq t < T} \big\|\bm Z^{N}[t]{-}\bm z[t]\big\|\geq \mu{-}\nu) \nonumber \\
\leq &  P_{\bm x}(\sup_{0 \leq t < T} \big\|\bm Z^{N}[t]{-}\bm z[t]\big\|\geq \mu{-}\nu) \nonumber \\
\leq &  \rho_1 \exp(-N \rho_2), \nonumber
\end{align}
which proves the lemma.
\end{proof}\vspace{5pt}

We let $\epsilon_s< \vartheta$ be such that if $\bm z[0] \in \Omega_{\epsilon_s}(\vec{\bm \zeta}^{\alpha}_{\bm
\gamma})$, then $\bm z[t] \in \Omega_{\epsilon}(\vec{\bm \zeta}^{\alpha}_{\bm
\gamma})$ for $t\geq 0$.

We let $\varrho^{N}_{2n}$, $n=0,1, \cdots$ be the time slots of \emph{consecutive} hitting times into the neighborhood $\Omega_{\epsilon_s}(\vec{\bm \zeta}^{\alpha}_{\bm
\gamma})$ from \emph{outside} of the neighborhood when the total number of users is $N$. Similarly, we let $\varrho^{N}_{2n+1}$, $n=0,1, \cdots$ denote the time slots of \emph{exiting} the
neighborhood $\Omega_{\epsilon}(\vec{\bm \zeta}^{\alpha}_{\bm
\gamma})$ from inside of the neighborhood, when the total number of users is $N$. Hence $\bm y_n=\bm Z^N[\varrho^{N}_{n}]$, $n=0,1, \cdots$ evolves as a Markov chain. In steady state,
\begin{align}
\label{eq:frac}
&P\big({\bm Z}^{N}[\infty] \notin
\Omega_{\epsilon}(\vec{\bm \zeta}^{\alpha}_{\bm \gamma})\big)\leq\frac{E[\varrho^{N}_{2n+2}-\varrho^{N}_{2n+1}]}{E[\varrho^{N}_{2n+2}-\varrho^{N}_{2n}]}\nonumber\\
=&\frac{E[\varrho^{N}_{2n+2}-\varrho^{N}_{2n+1}]}{E[\varrho^{N}_{2n+2}-\varrho^{N}_{2n+1}]+E[\varrho^{N}_{2n+1}-\varrho^{N}_{2n}]}.
\end{align}


We let $T_{\epsilon}(N)$ denote the random variable $\varrho^{N}_{2n+1}-\varrho^{N}_{2n}$. For any constant $K>0$, we have
\begin{align}
&E[T_{\epsilon}(N)]= \sum_{t=1}^{\infty} t \cdot P(T_{\epsilon}(N)=t) \nonumber \\
\geq& 2K \cdot P(T_{\epsilon}(N)\geq 2K)  \nonumber\\
=& 2K{\cdot}P_{\bm Z^N[\varrho^{N}_{2n+1}]}\Big(\sup_{\varrho^{N}_{2n+1} \leq t <\varrho^{N}_{2n+1+2K}}\hspace{-8pt}\| \bm Z^{N}[t]- \vec{\bm \zeta}^{\alpha}_{\bm
\gamma}\|\leq \epsilon  \Big). \label{eq:Kbound}
\end{align}

Note that
\begin{align}
&P_{\bm Z^N[\varrho^{N}_{2n+1}]}\Big(\sup_{\varrho^{N}_{2n+1} \leq t <\varrho^{N}_{2n+1}+2K} \| \bm Z^{N}[t]- \vec{\bm \zeta}^{\alpha}_{\bm
\gamma} \|> \epsilon  \Big) \nonumber \\
=& \hspace{-9pt}\sum_{\bm z \in \Omega_{\epsilon_s}(\vec{\bm \zeta}^{\alpha}_{\bm \gamma})} \hspace{-12pt}P\big(\bm Z^{N}(\varrho^{N}_1){=}\bm z\big) P_{\bm z}\Big(\sup_{0 \leq t < 2K} \| \bm Z^{N}[t]- \vec{\bm \zeta}^{\alpha}_{\bm
\gamma}\|{>}\epsilon\Big). \label{eq:meas1}
\end{align}

Since $\epsilon_s< \vartheta$, from Lemma~\ref{lemma:Kurtz_Index3}, there exist positive constants $\varsigma_1$ and $\varsigma_2$ such that for any $\bm z\in \Omega_{\epsilon_s}(\vec{\bm \zeta}^{\alpha}_{\bm \gamma})$,
\begin{align}
P_{\bm z}\Big(\sup_{0 \leq t < 2K} \| \bm Z^{N}[t]- \vec{\bm \zeta}^{\alpha}_{\bm
\gamma}\|> \epsilon  \Big) \leq \varsigma_1 \exp(-\varsigma_2 N). \label{eq:meas2}
\end{align}

Substitute (\ref{eq:meas2}) in (\ref{eq:meas1}) we have
\begin{align}
&P_{\bm Z^N[\varrho^{N}_{2n+1}]}\Big(\sup_{\varrho^{N}_{2n+1} \leq t <\varrho^{N}_{2n+1}+2K} \| \bm Z^{N}[t]- \vec{\bm \zeta}^{\alpha}_{\bm
\gamma} \|> \epsilon  \Big)\nonumber\\
\leq& \varsigma_1 \exp(-\varsigma_2 N). \nonumber
\end{align}

Therefore, $P_{\bm Z^{N_m}[\varrho^{N_m}_{2n+1}]}\Big(\sup_{\varrho^{N_m}_{2n+1} \leq t <\varrho^{N_m}_{2n+1+2K}} \| \bm Z^{N_m}[t]- \vec{\bm \zeta}^{\alpha}_{\bm
\gamma}\|\leq \epsilon  \Big)\rightarrow 1$ as $m \rightarrow \infty$. From (\ref{eq:Kbound}), if $m$ is large enough, we have
\begin{align}
E[T_{\epsilon}(N_m)]& \geq K. \nonumber
\end{align}

Since $K$ can be arbitrarily large, $\lim_{m \rightarrow \infty} E[T_{\epsilon}(N_m)]=\infty$, i.e., $\lim_{m \rightarrow \infty} E[\varrho^{N_m}_{2n+1}-\varrho^{N_m}_{2n}]=\infty$. Since from Assumption $\Psi$ we know $E[\varrho^{N_m}_{2n+2}-\varrho^{N_m}_{2n+1}]\leq M_{\epsilon_s}$, thus from equation~(\ref{eq:frac}),
\begin{align}
\lim_{m \rightarrow \infty} P\big({\bm Z}^{N_m}[\infty] \notin
\Omega_{\epsilon}(\vec{\bm \zeta}^{\alpha}_{\bm \gamma})\big)=0,
\nonumber
\end{align}
which concludes the proof.

\section{Proof of Proposition~\ref{prop:asymp}}
\label{appen:global}

For any $\ell>0$, let $\epsilon>0$ be such that for $\bm x \in \mathcal{Z}$, if $||\bm x -\bm \vec{\bm \zeta}^{\alpha}_{\bm \gamma}) ||<\epsilon$, then
\begin{align}
|v(\bm x)-r(\bm \gamma, \alpha)|<\ell. \nonumber
\end{align}

Consider fixed $N_m$, for $\forall \ell>0$ denote event $E_{N_m}=\{\bm Z^{N_m}[\infty] \in
\Omega_{\epsilon}(\vec{\bm \zeta}^{\alpha}_{\bm \gamma}) \}$, then
\begin{align}
&\Big | \frac{R^{N_m}_{\bm x}(\bm \gamma, \alpha)}{N_m}- r(\bm \gamma, \alpha)\Big | \nonumber \\
\leq& E\Big[ \big |v(Z^{N_m}[\infty])- v(\vec{\bm \zeta}^{\alpha}_{\bm \gamma})\big |  \Big] \nonumber \\
=& P\big(E_{N_m}\big) E\Big[\big| v(Z^{N_m}[\infty])- v(\vec{\bm \zeta}^{\alpha}_{\bm \gamma}) \big| \Big|  E_{N_m} \Big]\nonumber\\
&+ P\big(\bar{E}_{N_m}\big) E\Big[\big| v(Z^{N_m}[\infty])- v(\vec{\bm \zeta}^{\alpha}_{\bm \gamma}) \big| \Big|  \bar{E}_{N_m} \Big] \nonumber \\
\leq& P\big({\bm Z}^{N_m}[\infty] \in
\Omega_{\epsilon}(\vec{\bm \zeta}^{\alpha}_{\bm \gamma})\big) \cdot \ell + P\big({\bm Z}^{N_m}[\infty] \notin
\Omega_{\epsilon}(\vec{\bm \zeta}^{\alpha}_{\bm \gamma})\big). \label{eq:invar}
\end{align}

Apply Lemma~\ref{lemma:steady_dist} to (\ref{eq:invar}) we have
\begin{align}
&\lim_{m\rightarrow \infty} \Big | \frac{R^{N_m}_{\bm x}(\bm \gamma, \alpha)}{N_m}- r(\bm \gamma, \alpha)\Big |\nonumber\\
\leq& \lim_{m\rightarrow \infty} \Big[P\big({\bm Z}^{N_m}[\infty] \in
\Omega_{\epsilon}(\vec{\bm \zeta}^{\alpha}_{\bm \gamma})\big) {\cdot} \ell\nonumber\\
&\hspace{1.5in}{+} P\big({\bm Z}^{N_m}[\infty]{\notin}
\Omega_{\epsilon}(\vec{\bm \zeta}^{\alpha}_{\bm \gamma})\big)\Big]\nonumber\\
=&\ell. \nonumber
\end{align}

Since $\ell$ can be arbitrary,
\begin{align}
\lim_{m\rightarrow \infty} \frac{R^{N_m}_{\bm x}(\bm \gamma, \alpha)}{N_m}=r(\bm \gamma, \alpha) \nonumber,
\end{align}
which proves the proposition.

\section{Proof of Lemma~\ref{lemma:EigVal}}
\label{appen:EigVal}

After some algebra, the matrix $U^*$ takes the form
\begin{align}
\nonumber
U^* = \begin{bmatrix}\tilde{Q}^1(\bm z) &  B \\
0 & \tilde{Q}^2(\bm z)
\end{bmatrix}.
\end{align}
where matrix $B$ is expressed as
\begin{align}
\nonumber
B =\begin{bmatrix}
0&{\cdots}& 0 & b^1_{0,h_1^*}{-}1& b^1_{0,h_1^*}{-}1& {\cdots}& b^1_{0,h_1^*}{-}1\\
{\vdots} & & {\vdots} & & & &\\
0 & {\cdots} & 0 & 1 & 1 & {\cdots} &1\\
{\vdots}& & {\vdots} &  &  &  &\\
0& {\cdots} & 0 &  -b^1_{0,h_1^*}& -b^1_{0,h_1^*}& {\cdots}& -b^1_{0,h_1^*}\\
\end{bmatrix}
\end{align}
in which only the first, last and $h^*_1+1^{th}$ row have non-zero elements, and for each row, non-zero terms start at the $h^*_2$$^{th}$ element.

The matrices $\tilde{Q}^1(\bm z)$ and $\tilde{Q}^1(\bm z)$ are expressed in (\ref{eq:matrix1})(\ref{eq:matrix2}).

\begin{figure*}[hb]
\hrulefill
\begin{align}
\tilde{Q}^1(\bm z)&=\begin{bmatrix}
-1&0&\cdots& 0 & b^1_{0,h^*_1}-b^1_{0,h^*_1+1} & b^1_{0,h^*_1}-b^1_{0,h^*_1+2}  & \cdots& b^1_{0,h^*_1}-p_1 \\
1&-1& & & & & &\\
& \ddots & \ddots & & & & &\\
& &              1  & -1 & & & & \\
-1& \cdots     &    -1   & -1 & -1 & & &\\
& &                & &  & -1 & & \\
& &                & & & & \ddots &\\
& & &  & b^1_{0,h^*_1+1}-b^1_{0,h^*_1} & b^1_{0,h^*_1+1}-b^1_{0,h^*_1} & \cdots& -(1-p_1)-b^1_{0,h^*_1}
\end{bmatrix}\label{eq:matrix1}
\end{align}
\begin{align}
\tilde{Q}^2(\bm z)&=\begin{bmatrix}
-1&0&\cdots& 0 & 1{-} b^2_{0,h^*_2}& 1-b^2_{0,h^*_2+1} & \cdots& 1-p_2\\
1&-1& & & & & & &\\
& \ddots & \ddots & & & & &\\
& & 1  & -1 & & & &\\
-1& \cdots     &    -1   & -1 & -2 & -1 & \cdots &-1\\
& & & & &-1 & &\\
& & & & & & \ddots &\\
& & & & b^2_{0,h^*_2} & b^2_{0,h^*_2+1} & \cdots& -(1-p_2)
\end{bmatrix}.\label{eq:matrix2}
\end{align}
\vspace{-0.2in}
\end{figure*}

We need the following lemma to proceed.
\begin{lemma}
\label{lemma:bound}
For any $l \in \mathbb{Z}^+$,
\begin{align}
\nonumber
(1-p_1)+b^1_{0,l} > (l-1)(b^1_{0,l+1}-b^1_{0,l}).
\end{align}
\end{lemma}

\noindent \textbf{Proof:}
The proof is moved to Appendix~\ref{appen:bound}. $\hfill \blacksquare$
\vspace{3pt}

With this lemma, we proceed to characterize the eigen values of matrix $\bm U^*$, which are given by the solution to equation $\det(\bm U^*-\lambda I)=0$, where
\begin{align}
\det(\bm U^*-\lambda I )=\det \begin{bmatrix}\tilde{Q}^1(\bm z)-\lambda I &  B \\
& \tilde{Q}^2(\bm z)-\lambda I
\end{bmatrix}\nonumber\\
=\det\begin{bmatrix}\tilde{Q}^1(\bm z)-\lambda I & 0 \\
& \tilde{Q}^2(\bm z)-\lambda I
\end{bmatrix}, \nonumber
\end{align}
where the second equality is from the property of block matrices. Therefore, we have
\begin{align}
\nonumber
\det(\bm U^*-\lambda I )= \det (\tilde{Q}^1(\bm z)-\lambda I)\det(\tilde{Q}^2(\bm z)-\lambda I).
\end{align}

(1) We first study the characteristic polynomial $\det (\tilde{Q}^1(\bm z)-\lambda I)$. After some algebra we have
\begin{align}
&\det (\tilde{Q}^1(\bm z)-\lambda I)\nonumber\\
=&(1+\lambda)^{2\tau-h^*_1} \Big[[\lambda+(1{-}p_1){+}b_{0,h^*_1}^1](1+\lambda)^{h^*_1-1}\ - \nonumber \\ &(b_{0,h^*_1{+}1}^1{-}b_{0,h^*_1}^1)\big[1{+}(1{+}\lambda){+}(1{+}\lambda)^2{+}\cdots{+}(1{+}\lambda)^{h^*_1-2} \big]\Big] \nonumber \\
&\triangleq (1+\lambda)^{2\tau-h^*_1} \chi_1(\lambda). \nonumber
\end{align}
where
\begin{align}
\nonumber
&\chi_1(\lambda)=[\lambda{+}(1{-}p_1){+}b_{0,h^*_1}^1](1{+}\lambda)^{h^*_1{-}1}{-} (b_{0,h^*_1{+}1}^1{-}b_{0,h^*_1}^1)\nonumber\\
&\hspace{0.3in}\cdot\big[1{+}(1{+}\lambda){+}(1{+}\lambda)^2{+}\cdots{+}(1{+}\lambda)^{h^*_1{-}2} \big].
\end{align}

The matrix $\tilde{Q}^1(\bm z)$ hence has eigen value $-1$ of multiplicity $2\tau-h^*_1$. Let $\lambda$ be any other eigen value of $\tilde{Q}^1(\bm z)$, we hence have , i.e., $\chi_1(\lambda)=0$, i.e.,
\begin{align}
\label{eq:g01}
&[\lambda{+}(1{-}p_1){+}b_{0,h^*_1}^1](1{+}\lambda)^{h^*_1{-}1}= (b_{0,h^*_1{+}1}^1{-}b_{0,h^*_1}^1)\nonumber\\
&\cdot\big[1{+}(1{+}\lambda){+}(1{+}\lambda)^2{+}\cdots{+}(1{+}\lambda)^{h^*_1{-}2} \big].
\end{align}

We proceed to show that $\big|\lambda+1\big|<1$. We prove this by contradiction, suppose $\lambda$ is such that $\big|\lambda+1\big|\geq 1$. Then taking modulus of the left hand side of equation (\ref{eq:g01}) we have
\begin{align}
\big| &[\lambda{+}(1{-}p_1){+}b_{0,h^*_1}^1](1{+}\lambda)^{h^*_1{-}1}\big|\nonumber\\
=& \big|(\lambda+1)-(p_1-b^1_{0,h^*_1})\big| \cdot \big| 1+\lambda\big|^{h^*_1-1} \nonumber \\
\geq& \big||\lambda+1|-|p_1-b^1_{0,h^*_1}|\big|\cdot \big| 1+\lambda\big|^{h^*_1-1} \nonumber \\
\geq& \Big(1 - p_1 + b^1_{0,h^*_1}\Big) \big| 1+\lambda\big|^{h^*_1-1}, \nonumber
\end{align}
where the first equality is from triangle inequality. Applying Lemma~\ref{lemma:bound} we have,
\begin{align}
&\Big(1 - p_1 + b^1_{0,h^*_1}\Big) \big| 1+\lambda\big|^{h^*_1-1} \nonumber \\
>&(h^*_1-1)(b^1_{0,h^*_1+1}-b^1_{0,h^*_1}) \cdot \big| 1+\lambda\big|^{h^*_1-1} \nonumber \\
>&(b^1_{0,h^*_1+1}-b^1_{0,h^*_1})\big[1+\big| 1+\lambda\big|+\cdots+\big| 1+\lambda\big|^{h^*_1-2}\big] \nonumber \\
\geq & (b^1_{0,h^*_1+1}-b^1_{0,h^*_1}) \big| 1{+}(1{+}\lambda){+}\cdots{+}(1{+}\lambda)^{h^*_1{-}2} \big|. \label{eq:contra1}
\end{align}
where the first inequality is from Lemma~\ref{lemma:bound}, and the second inequality is from the fact that $\big|\lambda+1\big|{>}1$, and the last inequality comes from triangle Inequality. Note that inequality (\ref{eq:contra1}) contradicts (\ref{eq:g01}). Therefore each eigen values of matrix $\tilde{Q}^1(\bm z)$ must satisfy $\big|\lambda+1\big|< 1$.
\vspace{8pt}

(2) We then study the characteristic polynomial $\det (\tilde{Q}^2(\bm z)-\lambda I)$. We derive that
\begin{align}
&\det (\tilde{Q}^2(\bm z)-\lambda I) \nonumber\\
=& (1+\lambda)^{2\tau-h^*_2} \Big[\big[(1-p_2)+(1-b_{0, h^*_2}^2)\lambda \big]\nonumber\\
&\cdot\Big[1+(1+\lambda)+\cdots + (1+\lambda)^{h^*_2-3} \Big]\nonumber \\
& +(1+\lambda)^{h^*_2-2}\Big[\big[(1-p_2)+\lambda \big](2+\lambda)+ b_{0, h^*_2}^2 \Big] \Big] \nonumber \\
\triangleq & (1+\lambda)^{2\tau-h^*_2} \cdot \chi_2(\lambda), \label{eq:Q2zero}
\end{align}
where
\begin{align}
&\chi_2(\lambda)\nonumber\\
{=}&\big[(1{-}p_2){+}(1{-}b_{0, h^*_2}^2)\lambda \big]\Big[1{+}(1{+}\lambda){+}\cdots {+} (1{+}\lambda)^{h^*_2{-}3} \Big] \nonumber\\
&\hspace{0.3in}{+}(1{+}\lambda)^{h^*_2{-}2}\cdot\Big[\big[(1{-}p_2){+}\lambda \big](2{+}\lambda){+} b_{0, h^*_2}^2\Big].\nonumber
\end{align}

Consider
\begin{align}
&\lambda \cdot \chi_2(\lambda)\nonumber\\
{=}&\big[(1{-}p_2){+}(1{-}b_{0, h^*_2}^2)\lambda \big]\lambda \Big[1{+}(1{+}\lambda){+}\cdots {+} (1{+}\lambda)^{h^*_2{-}3} \Big]\nonumber\\
&\hspace{0.5in} {+} (1{+}\lambda)^{h^*_2{-}2} \lambda \Big[\big[(1{-}p_2){+}\lambda \big](2{+}\lambda){+} b_{0, h^*_2}^2 \Big]\nonumber \\
{=}&\big[(1{-}p_2){+}(1{-}b_{0, h^*_2}^2)\lambda \big](1{+}\lambda{-}1) \Big[1{+}(1{+}\lambda){+}\cdots\nonumber\\
&{+}(1{+}\lambda)^{h^*_2{-}3} \Big]\nonumber\\
&{+}(1{+}\lambda)^{h^*_2{-}2} \lambda \Big[\big[(1{-}p_2){+}\lambda \big](2{+}\lambda){+} b_{0, h^*_2}^2\Big] \nonumber \\
{=}&\big[(1{-}p_2){+}(1{-}b_{0, h^*_2}^2)\lambda \big]\big[(1{+}\lambda)^{h^*_2-2}-1\big]\nonumber\\
&\hspace{0.5in} {+}(1{+}\lambda)^{h^*_2{-}2} \lambda\Big[ \big[(1{-}p_2){+}\lambda \big](2{+}\lambda){+} b_{0, h^*_2}^2 \Big]\nonumber \\
{=}&-\big[(1{-}p_2){+}(1{-}b_{0, h^*_2}^2)\lambda \big]{+}(1{+}\lambda)^{h^*_2{-}2}\Big[\lambda \big[(1{-}p_2)\nonumber\\
&{+}\lambda \big](2{+}\lambda){+}b_{0, h^*_2}^2 \lambda{+}\big[(1{-}p_2){+}(1{-}b_{0, h^*_2}^2)\lambda \big] \Big]\nonumber \\
{=}&-\big[(1{-}p_2){+}(1{-}b_{0, h^*_2}^2)\lambda \big] {+}(1{+}\lambda)^{h^*_2{-}2}\nonumber\\
&\hspace{0.8in}\cdot\Big[\lambda\Big[ \big[(1{-}p_2){+}\lambda \big](2{+}\lambda){+} 1 \Big]+(1{-}p_2)\Big]\nonumber \\
&{=}-\big[(1{-}p_2){+}(1{-}b_{0, h^*_2}^2)\lambda \big]\nonumber\\
&{+}(1{+}\lambda)^{h^*_2{-}2} \Big[\lambda\Big[ (1{-}p_2)(2{+}\lambda){+} (\lambda+1)^2 \Big]+(1{-}p_2)\Big]\nonumber \\
&{=}-\big[(1{-}p_2){+}(1{-}b_{0, h^*_2}^2)\lambda \big]\hspace{1in}\nonumber\\
&\hspace{0.6in}{+}(1{+}\lambda)^{h^*_2{-}2} \Big[ (1{-}p_2)(1{+}\lambda)^2{+} \lambda(\lambda+1)^2 )\Big]\nonumber \\
&{=}-\big[(1{-}p_2){+}(1{-}b_{0, h^*_2}^2)\lambda \big]\nonumber\\
&\hspace{1.1in} {+}(1{+}\lambda)^{h^*_2{-}2} \Big[(1{-}p_2{+}\lambda)(\lambda+1)^2 \Big] \nonumber \\
&{=}{-}\big[(1{-}p_2){+}(1{-}b_{0, h^*_2}^2)\lambda \big] {+}(1{+}\lambda)^{h^*_2} (1{-}p_2{+}\lambda).\label{eq:lambdah}
\end{align}

It is clear from equation (\ref{eq:Q2zero}) that matrix $\tilde{Q}^2(\bm z)$ has eigen value $-1$ with multiplicity $2\tau-h^*_2$. Let $\lambda$ be any eigen value of $\tilde{Q}^2(\bm z)$, we first show the following lemma.

\begin{lemma}
\label{lemma:real_bound}
Let $\lambda$ be any eigen value of $\tilde{Q}^2(\bm z)$, then $-2<Re(\lambda)<0$.
\end{lemma}

\begin{proof}
1) Suppose $\tilde{Q}^2(\bm z)$ has an eigen value of $0$, then, from (\ref{eq:Q2zero}), $\chi_2(0)=0$. However,
\begin{align}
\chi_2(0)&= (1{-}p_2)(h^*_2{-}2){+} 2(1{-}p_2) {+} b_{0, h^*_2}^2 \nonumber \\
&= h^*_2 (1{-}p_2) {+} b_{0, h^*_2}^2 \nonumber \\
& \neq 0, \nonumber
\end{align}
leading to a contradiction. Hence $\tilde{Q}^2(\bm z)$ does not have $0$ eigen value.

2) Suppose the equation $\chi_2(\lambda)=0$ has a root $\lambda^*=a+b i$ with $a>0$, or $a\leq-2$, or being purely imaginary with $a=0, b \neq 0$. Hence from equation (\ref{eq:lambdah}),
\begin{align}
(1{-}p_2){+}(1{-}b_{0, h^*_2}^2)\lambda^* {=}(1{+}\lambda^*)^{h^*_2} \label{eq:x_xi2} (1{-}p_2{+}\lambda^*).
\end{align}

Consider the  modulus of the right hand side,
\begin{align}
&|(1{+}a+b i)^{h^*_2}| \cdot | 1-p_2+a+b i | \nonumber\\
>& | 1-p_2+ a+b i | \nonumber \\
>& | 1-p_2+(1{-}b_{0, h^*_2}^2)(a+b i) | \nonumber \\
=& | 1-p_2+(1{-}b_{0, h^*_2}^2)\lambda^* |. \nonumber
\end{align}

The above expression contradicts the previous equation (\ref{eq:x_xi2}).
\vspace{4pt}

From 1) and 2) we conclude that $\chi_2(\lambda)=0$ can only have solution with real part within $(-2,0)$. Therefore all eigen values of matrix $\tilde{Q}^2(\bm z)$ have real part within $(-2,0)$.
\end{proof}
\vspace{5pt}

We proceed to show that each eigen value $\lambda$ of $\tilde{Q}^2(\bm z)$ needs to satisfy $\big| \lambda+1 \big|<1$.
\vspace{3pt}

Suppose the equation $\chi_2(\lambda)=0$ has a root $\lambda$ with $\big|\lambda+1\big|\geq 1$, then from equation (\ref{eq:lambdah}),
\begin{align}
\label{eq:x_xi3}
(1{-}p_2){+}(1{-}b_{0, h^*_2}^2)\lambda {=}(1{+}\lambda)^{h^*_2} (1{-}p_2{+}\lambda).
\end{align}

We let $1+\lambda=x+yi$ where $x,y \in \mathbb{R}$. From the previous lemma we know that $|x|<1$. Some derivation shows that
\begin{align}
&|(1{-}p_2{+}\lambda)|^2 - |(1{-}p_2){+}(1{-}b_{0, h^*_2}^2)\lambda|^2 \nonumber \\
=& |1+\lambda|^2(2-b_{0, h^*_2}^2)b_{0, h^*_2}^2-2 x b_{0, h^*_2}^2 (1-p_2-b_{0, h^*_2}^2)\nonumber\\
&\hspace{2in}+b_{0, h^*_2}^2 (2p_2-b_{0, h^*_2}^2) \nonumber \\
> & |x| (2-b_{0, h^*_2}^2)b_{0, h^*_2}^2-2 |x| b_{0, h^*_2}^2 (1-p_2-b_{0, h^*_2}^2)\nonumber\\
&\hspace{1.8in}+|x| b_{0, h^*_2}^2 (2p_2-b_{0, h^*_2}^2) \nonumber \\
=& |x|b_{0, h^*_2}^2\Big[ (2{-}b_{0, h^*_2}^2){-}2 (1{-}p_2{-}b_{0, h^*_2}^2) +(2p_2{-}b_{0, h^*_2}^2) \Big] \nonumber \\
=& 0. \nonumber
\end{align}
where the first inequality is from the assumption that $|1+\lambda|\geq 1$ and the fact that $|x|<1$. Therefore
\begin{align}
|(1{-}p_2{+}\lambda) (1+\lambda)^{h^*_2} | & \geq |(1{-}p_2{+}\lambda) | \nonumber \\
&> |(1{-}p_2){+}(1{-}b_{0, h^*_2}^2)\lambda|. \nonumber
\end{align}

The above expression contradicts equation (\ref{eq:x_xi3}). Hence it can not be $\big|\lambda+1\big|\geq 1$. Therefore, each eigen value $\lambda$ of $\bm U^*$ satisfies $\big|\lambda+1\big|< 1$, which concluds the proof.

\section{Proof of Proposition~\ref{prop:discrete_kurtz}}
\label{appen:discrete_kurtz}
Consider the random variable $\bm Z^{N}[t+1]$ given $\bm Z^{N}[t]=\bm z$, i.e.,
\begin{align}
\bm Z^{N}[t+1]{=}\bm Z^{N}[t]+\sum_{i,j=1}^{2(2\tau+1)} \frac{\sum_{h=1}^{N z_i} \eta^h_{ij}(\bm z) {\cdot} \bm e_{ij}}{N}, \label{eq:drift_t}
\end{align}
where $\eta^h_{ij}(\bm z)$ is an indicator function representing whether the belief value of the $h^{th}$ user transits from belief value $\beta_i$ to belief value $\beta_j$ at the next time slot. Note that, given $\bm Z^{N}[t]=\bm z$, the scheduling action for users in belief state $\beta_i$ is independent of $N$ because the scheduling decision only depends on the belief state distribution $\bm z$. As $N$ increases and $\bm z$ stays unchanged, more users are in belief state $\beta_i$ and the contribution of each channel to the transition of $\bm Z^{N}$ scales down with $N$. From the law of large numbers, if the number of users scales up while $z_i$ is kept the same, we have
\begin{align}
\lim_{N \rightarrow \infty} \frac{\sum_{h=1}^{N z_i} \eta^h_{ij}(\bm z)}{N}{=}&\lim_{N \rightarrow \infty} \frac{N z_i}{N} \frac{\sum_{h=1}^{N z_i} \eta^h_{ij}(\bm z)}{N z_i}{=}z_i q_{ij}(\bm z)\nonumber\
\end{align}
almost surely, where $q_{ij}(\bm z)$ is defined in~(\ref{eq:qz}).

\begin{lemma}
\label{lemma:onestep}
There exists a neighborhood $\Omega_{\varepsilon}(\vec{\bm \zeta})$ of $\vec{\bm \zeta}$ such that, for any $\mu>0$, if $Z^{N}[t]=\bm z \in \Omega_{\varepsilon}(\bm \zeta)$, there exists a function $f(\mu)$ for which $Z^{N}[t+1]$ satisfies
\begin{align}
&P\Big( \big\| \bm Z^{N}[t+1]-\big(I+Q(\bm z)\big)\bm z \big\| \geq \mu \Big| Z^{N}[t]=\bm z \Big)\nonumber\\
\leq& 4 \exp(-N \cdot f(\mu)), \nonumber
\end{align}
where $f(\mu)$ is independent of $\bm z$ and $N$.
\end{lemma}

\begin{proof}
Let $\vec{\bm 1}_i$ be a vector with $1$ at the $i^{th}$ position and $0$ in all other elements. From (\ref{eq:drift_t}),
\begin{align}
&\bm Z^{N}[t+1]-\big(I+Q(\bm z)\big)\bm z \nonumber \\
=& \sum_{i,j=1}^{2(2\tau+1)} \frac{\sum_{h=1}^{N z_i} \eta^h_{ij}(\bm z)}{N} \cdot \bm e_{ij}- Q(\bm z)\bm z \nonumber \\
=& \sum_{i,j=1}^{2(2\tau+1)} \frac{\sum_{h=1}^{N z_i} \eta^h_{ij}(\bm z)}{N} \cdot \bm e_{ij}- \sum_{i,j=1}^{2(2\tau+1)}z_i q_{ij}(\bm z)\cdot \bm e_{ij} \nonumber \\
=& \sum_{i,j=1}^{2(2\tau+1)} \frac{\sum_{h=1}^{N z_i} \eta^h_{ij}(\bm z)}{N} \cdot \big(\vec{\bm 1}_{j}-\vec{\bm 1}_{i}\big)\nonumber\\
&\hspace{1.2in} - \sum_{i,j=1}^{2(2\tau+1)}z_i q_{ij}(\bm z)\cdot \big(\vec{\bm 1}_{j}-\vec{\bm 1}_{i}\big) \nonumber \\
=& \Big[ \sum_{i,j=1}^{2(2\tau+1)} \frac{\sum_{h=1}^{N z_i} \eta^h_{ij}(\bm z)}{N} \cdot \vec{\bm 1}_{j} - \sum_{i,j=1}^{2(2\tau+1)}z_i q_{ij}(\bm z)\cdot \vec{\bm 1}_{j}\Big]\nonumber\\
&-\Big[ \sum_{i,j=1}^{2(2\tau+1)} \frac{\sum_{h=1}^{N z_i} \eta^h_{ij}(\bm z)}{N} \cdot \vec{\bm 1}_{i} {-} \sum_{i,j=1}^{2(2\tau+1)}z_i q_{ij}(\bm z)\cdot \vec{\bm 1}_{i}\Big]. \nonumber
\end{align}

Note that
\begin{align}
&\sum_{i,j=1}^{2(2\tau+1)} \frac{\sum_{h=1}^{N z_i} \eta^h_{ij}(\bm z)}{N} {\cdot} \vec{\bm 1}_{i} {-} \sum_{i,j=1}^{2(2\tau+1)}z_i q_{ij}(\bm z){\cdot} \vec{\bm 1}_{i}\nonumber\\
{=}&\sum_{i=1}^{2(2\tau+1)} \frac{\sum_{h=1}^{N z_i} \sum_{j=1}^{2(2\tau+1)} \eta^h_{ij}(\bm z)}{N} {\cdot} \vec{\bm 1}_{i}\nonumber\\
&\hspace{1in}{-} \sum_{i=1}^{2(2\tau+1)}z_i \sum_{j=1}^{2(2\tau+1)} q_{ij}(\bm z){\cdot} \vec{\bm 1}_{i} \nonumber \\
{=}&\sum_{i=1}^{2(2\tau+1)} z_i \vec{\bm 1}_{i} -\sum_{i=1}^{2(2\tau+1)} z_i \vec{\bm 1}_{i} \nonumber \\
=&0, \nonumber
\end{align}
where the second equality holds because $\sum_{j=1}^{2(2\tau+1)} \eta^h_{ij}(\bm z)=1$ for all $h$, and $\sum_{j=1}^{2(2\tau+1)} q_{ij}(\bm z)=1$ for all $i$.

Therefore
\begin{align}
&\bm Z^{N}[t+1]-\big(I+Q(\bm z)\big)\bm z \nonumber\\
=& \sum_{i,j=1}^{2(2\tau+1)} \frac{\sum_{h=1}^{N z_i} \big(\eta^h_{ij}(\bm z)-q_{ij}(\bm z)\big)}{N}\cdot \vec{\bm 1}_{j}. \label{eq:Zupdate}
\end{align}

Note that once a user is activated, its belief value will only transit to $p_k$ or $r_k$, therefore $\eta^h_{ij}(\bm z)\neq 0$ only for $j\in\Theta:=\{1, 2\tau+1, 2\tau+2, 2(2\tau+1)\}$. Also note that for those channels that stay idle, there is no randomness associated with its belief transition, i.e., for them $\eta^h_{ij}(\bm z)=q_{ij}(\bm z) \in \{0, 1\}$. Therefore the randomness is only associated with the channels which are activated, i.e., those with index value no smaller than $\omega^*$. Hence, (\ref{eq:Zupdate}) becomes
\begin{align}
&\bm Z^{N}[t+1]-\big(I+Q(\bm z)\big)\bm z \nonumber\\
=& \sum_{j\in \Theta}\sum_{i \in \Pi_j(\bm z)} \frac{\sum_{h=1}^{N g_i(\bm z) z_i} \big(\eta^h_{ij}(\bm z)-q_{ij}(\bm z)\big)}{N}\cdot \vec{\bm 1}_{j}, \nonumber
\end{align}
where the summation $\sum_{h=1}^{N g_i(\bm z) z_i}(\cdot)$ is over all the channels in belief state $\beta_i$ that are activated, and $\Pi_j(\bm z)$ is the set of belief values in which channels are scheduled within the class that corresponds to belief $j \in \Theta$, i.e.,
\begin{align}
\Pi_j(\bm z){:=}\begin{cases}
\{1\leq i \leq2\tau+1: g_i(\bm z) > 0 \} \text{if $j=1, 2\tau{+}1,$} \\
\{(2\tau+1)+1 \leq i \leq 2(2\tau+1): g_i(\bm z) > 0 \}\\
\hspace{1in}\text{if $j=2\tau+2, 2(2\tau+1).$}
\end{cases} \nonumber
\end{align}

We hence have
\begin{align}
&P\Big( \big\| \bm Z^{N}[t{+}1]{-}\big(I{+}Q(\bm z)\big)\bm z \big\| {\geq} \mu \Big| Z^{N}[t]{=}\bm z \Big)\nonumber\\
=& P\Big( \big\|\sum_{j\in \Theta}\sum_{i \in \Pi_j(\bm z)} \hspace{-5pt}\frac{\sum_{h=1}^{g_i(\bm z)N z_i} \big(\eta^h_{ij}(\bm z){-}q_{ij}(\bm z)\big)}{N}{\cdot} \vec{\bm 1}_{j}\big\| {>} \mu \Big) \nonumber \\
\leq & \sum_{j\in \Theta} P\Big( \Big|\sum_{i \in \Pi_j(\bm z)}\sum_{h=1}^{g_i(\bm z)N z_i}  \frac{ \eta^h_{ij}(\bm z){-}q_{ij}(\bm z)}{N}\Big| {>} \frac{\mu}{4} \Big), \label{eq:deviation}
\end{align}
where the last inequality holds because $\big|\Theta\big|=4$ as well as the union bound. Specifically, the union bound holds since
\begin{align}
&\big\{ \Big\|\sum_{j\in \Theta}\sum_{i \in \Pi_j(\bm z)} \frac{\sum_{h=1}^{g_i(\bm z)N z_i} \big(\eta^h_{ij}(\bm z){-}q_{ij}(\bm z)\big)}{N}{\cdot} \vec{\bm 1}_{j}\Big\| > \mu\big\}\nonumber\\
\subseteq& \bigcup_{j\in\Theta}\big\{\Big|\sum_{i \in \Pi_j(\bm z)}\sum_{h=1}^{g_i(\bm z)N z_i}  \frac{ \eta^h_{ij}(\bm z)-q_{ij}(\bm z)}{N}\Big| > \frac{\mu}{4} \big\}.\nonumber
\end{align}

Note that, for each $j\in \Theta$, the random variables $\eta^h_{ij}(\bm z), h=1,\cdots, g_i(\bm z)N z_i$ are independent. From an extension of Chebychoff's inequality (See Excercise 1.8 in \cite{Weiss_LD}) we have that, for each $j \in \Theta$, there exists a positive continuous function $f_{j}(\mu)$, which does not depend on $\bm z$ and $N$, with
\begin{align}
&P\Big( \Big|\sum_{i \in \Pi_j(\bm z)}\sum_{h=1}^{g_i(\bm z)N z_i} \frac{\eta^h_{ij}(\bm z)-q_{ij}(\bm z)}{N}\Big|> \frac{\mu}{4} \Big) \nonumber\\
<& \exp\big(-f_{j}(\mu) \sum_{i \in \Pi_j(\bm z)} g_i(\bm z)N z_i\big).
\end{align}

Let $\alpha_j$ be the fraction of channels activated, under the \emph{steady state} of Optimal Relaxed Policy, in the class corresponding to belief value $\beta_j$, i.e.,
\begin{align}
\alpha_j{=}\hspace{-3pt}\begin{cases}
\sum_{i=1}^{2\tau+1} g_i(\bm \zeta) \zeta_i &\hspace{-6pt}\text{if $j{\in}\{1, 2\tau{+}1\}$,} \\
\sum_{i=2\tau+2}^{2(2\tau+1)} g_i(\bm \zeta) \zeta_i &\hspace{-6pt}\text{if $j{\in}\{2\tau{+}2, 2(2\tau{+}1)\}$.}
\end{cases}
\end{align}

For any $0<\ell< \min\{\alpha_j, j \in \Theta \}$, there exists a neighborhood $\Omega_{\varepsilon}(\vec{\bm \zeta})$ such that for all $\bm z \in \Omega_{\varepsilon}(\vec{\bm \zeta})$,
\begin{align}
\sum_{i \in \Pi_j(\bm z)} g_i(\bm z)z_i \geq \alpha_j-\ell, \quad j\in \Theta, \label{eq:deviation2}
\end{align}
which essentially means, under system state $\bm z \in \Omega_{\varepsilon}(\vec{\bm \zeta})$, the fraction of activated channels in each class will stay close to the case when system state is actually $\bm \zeta$. Let $f(\mu)=\min\{f_{j}(\mu)(\alpha_j-\ell), j\in \Theta \}$, then from (\ref{eq:deviation})-(\ref{eq:deviation2}),
\begin{align}
&P\Big( \big\| \bm Z^{N}[t+1]-\big(I+Q(\bm z)\big)\bm z \big\| \geq \mu \Big| Z^{N}[t]=\bm z \Big)\nonumber\\
\leq &\sum_{j\in \Theta} P\Big( \Big|\sum_{i \in \Pi_j}\sum_{h=1}^{g_i(\bm z)N z_i}  \frac{ \eta^h_{ij}(\bm z)-q_{ij}(\bm z)}{N}\Big| > \frac{\mu}{4} \Big) \nonumber \\
\leq& 4\exp(-N\cdot f(\mu)).\nonumber
\end{align}

It is clear from the proof that $f(\mu)$ does not depend on $\bm z$ or $N$. The lemma thus holds.
\end{proof}
\vspace{5pt}

\begin{lemma}
\label{lemma:onestep2}
There exists a neighborhood $\Omega_{\delta}(\bm \zeta)$ of $\vec{\bm \zeta}$ such that, for any $\mu>0$, if $Z^{N}[0]=\bm x \in \Omega_{\delta}(\bm \zeta)$, for any $t\geq 1$, there exist positive constant $c_1^t$ and $c_2^t$ with
\begin{align}
P_{\bm x}\Big( \big\| \bm Z^{N}[t]-\bm z[t] \big\| \geq \mu \Big)\leq c_1^t \cdot \exp(-N \cdot c_2^t), \nonumber
\end{align}
where $c_1^t$ and $c_2^t$ is independent of $\bm x$ and $N$.
\end{lemma}

\begin{proof}
We let $\nu<\mu$. From Lemma~\ref{lemma:onestep}, there exists $\varepsilon$ such that if $\bm z \in \Omega_{\varepsilon}(\bm \zeta)$
\begin{align}
&P\Big( \big\| \bm Z^{N}[t+1]-\big(I+Q(\bm z)\big)\bm z \big\| \geq \mu \Big| Z^{N}[t]=\bm z\Big)\nonumber\\
\leq& 4 \exp(-f(\mu)\cdot N). \label{eq:mu_nu_bound}
\end{align}

We let $\rho < \varepsilon$ be such that
\begin{align}
\label{eq:rho_bound}
\big\|\big(Q(\bm x)+I\big)\bm x - \big(Q(\bm y)+I\big) \bm y \big\| \leq \nu
\end{align}
for all $\bm x, \bm y \in \mathcal{Z}$ with $\big\|\bm x - \bm y\big\| \leq \rho.$
Recall that $\sigma$ is defined in Lemma~\ref{prof:fluid_conv}. We let $\delta<\min\{\sigma, \varepsilon \}$ be such that, if $\bm z[0] \in \Omega_{\delta}(\bm \zeta)$, $\bm z[t] \in \Omega_{\varepsilon-\rho}(\bm \zeta)$ for all $t\geq 1$. We proceed to prove this statement by induction.
\vspace{3pt}

For $t=1$, if $\bm x \in \Omega_{\delta}(\bm \zeta)$, from inequality~(\ref{eq:mu_nu_bound}), there exists $f(\mu)$,
\begin{align}
\nonumber
&P_{\bm x}(\big\| \bm Z^{N}[1]-\bm z[1] \big\| \geq \mu \Big)\nonumber\\
=&P_{\bm x}(\big\| \bm Z^{N}[1]-\big(I+Q(\bm x)\big)\bm x \big\| \geq \mu \Big)\nonumber\\
\leq& 4\exp(-f(\mu)\cdot N).
\end{align}

Letting $c_1^1=4$ and $c_1^2=f(\mu)$, the statement holds when $t=1$.

Suppose the statement is true at $t\geq 1$, then there exist $d_1^t$ and $d_2^t$, which correspond to $\rho$, for which,
\begin{align}
&P_{\bm x}\Big(\big\| \bm Z^{N}[t+1]-\bm z[t+1] \big\| \geq \mu \Big)\nonumber \\
=&P_{\bm x}\Big(\big\| \bm Z^{N}[t]-\bm z[t] \big\| \geq \rho \Big)\nonumber\\
&{\cdot} P_{\bm x}\Big(\big\| \bm Z^{N}[t{+}1]{-}\bm z[t{+}1] \big\| \geq \mu \Big| \big\| \bm Z^{N}[t]{-}\bm z[t] \big\|{\geq}\rho \Big) \nonumber \\
&+ P_{\bm x}\Big(\big\| \bm Z^{N}[t]-\bm z[t] \big\| < \rho \Big)\nonumber\\
&\hspace{-8pt}\cdot P_{\bm x}\Big(\big\| \bm Z^{N}[t{+}1]{-}\bm z[t{+}1] \big\| \geq \mu \Big| \big\| \bm Z^{N}[t]{-}\bm z[t] \big\| {<} \rho \Big) \nonumber \\
\leq& d_1^t\exp(-N\cdot d_2^t)\nonumber\\
&\hspace{-9pt}{+}P_{\bm x}\Big(\hspace{-2pt}\big\|\hspace{-2pt}\bm Z^{N}[t{+}1]{-}\bm z[t{+}1] \big\| {\geq} \mu \Big|\hspace{-2pt}\big\| \hspace{-2pt}\bm Z^{N}[t]{-}\bm z[t] \big\| {<} \rho \Big).\hspace{-3pt}\label{eq:t2_1}
\end{align}

Now consider the second term in (\ref{eq:t2_1}),
\begin{align}
&P_{\bm x}\Big(\big\|\bm Z^{N}[t{+}1]{-}\bm z[t{+}1] \big\| {\geq} \mu \Big| \ \big\| \bm Z^{N}[t]{-}\bm z[t] \big\| {<} \rho \Big) \nonumber \\
=& P_{\bm x}\Big(\hspace{-3pt}\big\|\hspace{-2pt}\bm Z^{N}[t{+}1]{-}\hspace{-1pt}\big(I{+}Q(\bm Z^{N}[t])\big) \bm Z^{N}[t]{+}\big(I{+}Q(\bm Z^{N}[t])\big)\nonumber\\
& \cdot\bm Z^{N}[t] -\bm z[t+1] \big\| \geq \mu \Big|\big\| \bm Z^{N}[t]-\bm z[t] \big\| < \rho \Big) \nonumber \\
\leq & P_{\bm x}\Big(\big\| \bm Z^{N}[t{+}1]{-}\big(I{+}Q(\bm Z^{N}[t])\big) \bm Z^{N}[t] \big\|{+}\nonumber\\
&\big\| \big(I{+}Q(\bm Z^{N}[t])\big) \bm Z^{N}[t]{-}\big(I{+}Q(\bm z[t])\big)\bm z[t] \big\| \geq \mu \nonumber\\
&\Big| \ \big\| \bm Z^{N}[t]{-}\bm z[t] \big\| {<} \rho \Big) \nonumber \\
\leq & P_{\bm x}\Big(\big\|\bm Z^{N}[t{+}1]{-}\big(I{+}Q(\bm Z^{N}[t])\big) \bm Z^{N}[t] \big\|\nonumber\\
&\hspace{1.4in} \geq \mu{-}\nu \Big| \ \big\| \bm Z^{N}[t]{-}\bm z[t] \big\| {<} \rho \Big) \nonumber \\
= & \sum_{\bm z \in \Omega_{\rho}(\bm z[t])} P_{\bm x}\big(\bm Z^{N}[t]{=}\bm z\Big| \bm Z^{N}[t] \in \Omega_{\rho}(\bm z[t]) \big)\nonumber\\
&P_{\bm x}\Big(\big\| \bm Z^{N}[t+1]{-} \big(I{+}Q(\bm z)\big) \bm z \big\|\geq \mu{-}\nu \Big| \ \bm Z^{N}[t]{=}\bm z \Big) \label{eq:t2_2ineq}
\end{align}
where the first inequality follows from triangle inequality, and the second inequality is from relationship (\ref{eq:rho_bound}).

Because $\bm z[t] \in \Omega_{\varepsilon-\mu}(\bm \zeta)$ and $\rho<\varepsilon$, we have $\Omega_{\rho}(\bm z[t]) \subseteq \Omega_{\varepsilon}(\bm \zeta)$. From inequality~(\ref{eq:mu_nu_bound}), we have
\begin{align}
&P_{\bm x}\Big(\big\| \bm Z^{N}[t{+}1]{-}\big(I{+}Q(\bm z)\big) \bm z \big\| {\geq} \mu{-}\nu \Big| \bm Z^{N}[t]=\bm z \Big)\nonumber\\
\leq& 4 \exp(-N\cdot f(\mu-\nu)). \label{eq:t2_2}
\end{align}

Substituting (\ref{eq:t2_2}) to (\ref{eq:t2_2ineq}), we have
\begin{align}
&P_{\bm x}\Big(\big\| \bm Z^{N}[t{+}1]{-}\bm z[t{+}1] \big\| {\geq}\mu \Big| \ \big\| \bm Z^{N}[t]-\bm z[t] \big\| < \rho \Big)\nonumber\\
\leq& 4 \exp(-N \cdot f(\mu-\nu)). \label{eq:t2_3}
\end{align}

Hence from Equation~(\ref{eq:t2_1}) and (\ref{eq:t2_3}), there exists constants $c_1^{t+1}>0$ and $c_2^{t+1}>0$ that do not depend on $\bm z$ and $N$ with
\begin{align}
P_{\bm x}\Big(\big\| \bm Z^{N}[t{+}1]{-}\bm z[t{+}1] \big\| {\geq} \mu \Big) \leq c_1^{t{+}1} \exp(-N c_2^{t{+}1}). \nonumber
\end{align}

By induction, the lemma holds.
\end{proof}
\vspace{3pt}

Note that from union bound,
\begin{align}
&P_{\bm x}\Big( \sup_{0 \leq t < T} || \bm Z^{N}[t]-\bm z[t] || \geq \mu \Big) \nonumber\\
\leq& \sum_{t=0}^{T-1} P_{\bm x}\Big(|| \bm Z^{N}[t]-\bm z[t] || \geq \mu \Big). \label{eq:Pdecomp2}
\end{align}

Therefore, from Lemma~\ref{lemma:onestep2}, over finite time horizon $T$, there exist positive constants $C_1$ and $C_2$, which do not depend on $\bm x$ and $N$, such that
\begin{align}
\nonumber
P_{\bm x}\Big( \sup_{0 \leq t < T} || \bm Z^{N}[t]-\bm z[t] || \geq \mu \Big) \leq C_1 \exp(-N C_2),
\end{align}
which concludes the proof of Proposition~\ref{prop:discrete_kurtz}.

\vspace{5pt}

\section{Proof of Lemma~\ref{lemma:bound}}
\label{appen:bound}

\begin{proof}
From the belief value evolution (\ref{eq:evolve}) we know
\begin{align}
b^1_{0,l}&=\frac{r_1-r_1(p_1-r_1)^l}{1+r_1-p_1}, & b^1_{0,l+1}-b^1_{0,l}&=r_1(p_1-r_1)^l. \nonumber
\end{align}

Therefore
\begin{align}
&(1-p_1)+\pi^1_{0,l} - (l-1)(\pi^1_{0,l+1}-\pi^1_{0,l})\nonumber \\
=&(1-p_1)+\frac{r_1-r_1(p_1-r_1)^{l}}{1+r_1-p_1} - (l-1)r_1(p_1-r_1)^{l} \nonumber \\
=&(1-p_1)+\frac{r_1-r_1(p_1-r_1)^{l}}{1+r_1-p_1} - (l-1)r_1(p_1-r_1)^{l} \nonumber \\
=&(1-p_1)+r_1\Big[ \frac{1-(p_1-r_1)^{l}}{1+r_1-p_1} - (l-1)(p_1-r_1)^{l} \Big] \nonumber \\
=&(1{-}p_1){+}r_1\Big[ \frac{1+(l-1)(p_1-r_1)^{l+1}-l(p_1-r_1)^{l}}{1+r_1-p_1} \Big] \nonumber\\
=&(1{-}p_1){+}r_1\Big[ \frac{1{+}(l{-}1)(p_1{-}r_1)^{l}(p_1{-}r_1{-}1){-}(p_1{-}r_1)^{l}}{1{+}r_1{-}p_1} \Big] \nonumber \\
=&(1{-}p_1){+}\frac{r_1}{1+r_1-p_1}\Big[ (l{-}1)(p_1{-}r_1)^{l}(p_1{-}r_1{-}1)\nonumber\\
&\hspace{0.1in}{-}(p_1{-}r_1{-}1)(1{+}(p_1{-}r_1)+\cdots+(p_1-r_1)^{l-1}) \Big] \nonumber \\
=&(1-p_1)+r_1\Big[ (1+(p_1-r_1)+\cdots+(p_1-r_1)^{l-1})\nonumber\\
&\hspace{1in}-(l-1)(p_1-r_1)^{l} \Big]
 \label{eq:lemma_diff}
\end{align}

Since $(p_1-r_1)^j \geq (p_1-r_1)^{l}$ for $l = 1, \cdots, j-1$, therefore from equation (\ref{eq:lemma_diff}),
\begin{align}
&(1-p_1)+\pi^1_{0,l} - (l-1)(\pi^1_{0,l+1}-\pi^1_{0,l})\nonumber\\
\geq& (1-p_1)+r_1 > 0, \nonumber
\end{align}
which proves the lemma.
\end{proof}

\end{document}